\documentclass{iopbk2e}
\usepackage[english]{babel}
\usepackage{iopams}
\usepackage{epsf}
\usepackage{epsfig}
\newcommand{\msunabel}{M_{\odot}}
\def\etal{{\it et al.~}}
\def\ie{{\frenchspacing\it i.e. }}
\def\eg{{\frenchspacing\it e.g. }}

\def\gtsima{$\; \buildrel > \over \sim \;$}
\def\ltsima{$\; \buildrel < \over \sim \;$}
\def\prosima{$\; \buildrel \propto \over \sim \;$}
\def\gsim{\lower.5ex\hbox{\gtsima}}
\def\lsim{\lower.5ex\hbox{\ltsima}}
\def\simgt{\lower.5ex\hbox{\gtsima}}
\def\simlt{\lower.5ex\hbox{\ltsima}}
\def\simpr{\lower.5ex\hbox{\prosima}}

\def\HH{H$_2$~}
\def\HHP{H$_2^+$}

\begin{document}
\Chapter{The Formation of Primordial Luminous Objects}
%
{Emanuele Ripamonti \\
Kapteyn Astronomical Institute, University of Groningen \\
The Netherlands \\~\\
Tom Abel \\
Kavli Institute for
Astroparticle Physics and Cosmology, Stanford University \\
U.S.A.}


\section{Introduction}

The scientific belief that the universe evolves in time is one of the legacies of the theory of the Big Bang. 

The concept that the universe has an history started to attract the interest of cosmologists soon after the first formulation of the theory: already Gamow (1948; 1949) investigated how and when galaxies could have been formed in the context of the expanding Universe.

However, the specific topic of the formation (and of the fate) of the {\it first} objects dates to two decades later, when no objects with metallicities as low as those predicted by primordial nucleosynthesis ($Z\lsim10^{-10}\sim10^{-8}Z_\odot$) were found.  Such concerns were addressed in two seminal papers by Peebles \& Dicke (1968; hereafter PD68) and by
Doroshkevich, Zel'Dovich \& Novikov (1967; hereafter DZN67)\footnote{This paper is in Russian and we base our comments on indirect knowledge (\eg from the Novikov \& Zel'Dovich 1967 review).}, introducing the idea that some objects could have formed before the stars we presently observe. 
\begin{enumerate}
\item{Both PD68 and DZN67 suggest a mass of $\sim10^5\msunabel$ for the first
generation of bound systems, based on the considerations on the cosmological
Jeans length\index{Jeans length} (Gamow 1948; Peebles 1965) and the possible shape of the power spectrum.}
\item{They point out the role of thermal instabilities in the formation of the
proto-galactic bound object, and of the cooling\index{cooling} of the gas
inside it; in particular, PD68 introduces \HH cooling\index{HH cooling} and
chemistry\index{HH chemistry} in the calculations about the contraction of the gas.}
\item{Even if they do not specifically address the occurrence of fragmentation\index{fragmentation}, these papers make two very different assumptions: PD68 assumes that the gas will fragment into ``normal'' stars to form globular clusters, while DZN67 assumes that fragmentation {\it does not}\ occur, and that a single ``super-star'' forms.}
\item{Finally, some feedback\index{feedback} effects as considered (\eg Peebles \& Dicke considered the effects of supernovae).}
\end{enumerate}

Today most of the research focuses on the issues when fragmentation\index{fragmentation} may occur, what objects are formed and how they influence subsequent structure formation.

In these notes we will leave the discussion of feedback to lecture notes by
Ferrara \& Salvaterra and by Madau \& Haardt in this same book and focus only on the aspects of the formation of the first objects.

The advent of cosmological numerical hydrodynamics in particular allow a fresh new look at these questions. Hence, these notes will touch on aspects of theoretical cosmology to chemistry\index{chemistry}, computer science, hydrodynamics and atomic physics. 
For further reading and more references on the subject we refer the reader to other
relevant reviews such as Barkana \& Loeb 2001, and more recently Ciardi \& Ferrara 2004,
Glover 2004 and Bromm \& Larson 2004.

In this notes, we try to give a brief introduction to only the most relevant
aspects. We will start with a brief overview of the relevant cosmological
concepts in section 2, followed by a discussion of the properties of
primordial material (with particular emphasis to its cooling and its
chemistry\index{chemistry}) in section 3.  We will then review the technique
and the results of numerical simulations\index{numerical Simulation} in sections 4 and 5: the former will deal with detailed 3D simulations of the formation of gaseous clouds which are likely to transform into luminous objects, while the latter will examine results (mostly from 1D codes) about the modalities of such transformation. Finally, in section 6 we will critically discuss the results of the previous sections, examining their consequences and comparing them to our present knowledge of the universe.


\section{Physical cosmology}

In the following of this notes we will adopt the modern physical
description of the universe (dating back to at least 1920), based upon
the ``cosmological principle'', which affirms that on cosmological
scales the distributions of matter and energy should be homogeneous
and isotropic, whose metric is the Robertson-Walker metric. Although
derived from mainly philosophical arguments, the
cosmological principle is also supported by observations such as the
isotropy of the CMB ({\it e.g.} Wu \etal 1999).

We will also make some additional, general assumptions which are
quite common in present-day cosmology, and which are believed to be
the best explanations for a series of observations. That is:
\begin{enumerate}
\item{The cosmological structures we observe at present (galaxies,
clusters of galaxies etc.) formed because of gravitational instability
of pre-existing, much shallower fluctuations;}
\item{Most of the matter in the universe is in the form of {\it ``Cold
Dark Matter''}\index{dark matter}, that is, of some kind of elementary particle (or
particles) that has not been discovered at present. Cold Dark Matter\index{dark matter}
particles are assumed to have a negligibly small cross section for
electromagnetic interactions (\ie to be {\it dark}), and to move at
non-relativistic speeds (\ie to be{\it cold}).}
\end{enumerate}



\subsection{Fluctuations in the Early Universe}

\subsubsection{Inflation\index{inflation}}
Inflation is a mechanism which was first proposed by Guth (1981) and (in
a different way) by Starobinsky (1979, 1980), and has since been
included in a number of different ``inflationary theories'' (see Guth
2004 for a review upon which we base this paragraph).

The basic assumption of inflationary models is the existence of states
with negative pressure; a typical explanation is that some unidentified
kind of scalar field (commonly referred to as {\it inflaton}\index{inflation})
temporarily keeps part of the universe in a ``false vacuum'' state, in
which the energy density must be approximately constant (that is, it can
not ``diluted'' by expansion) at some value $\rho_f c^2$, implying a
negative pressure $p_f=-\rho_f c^2$. Inserting these two expression in
the first Friedmann cosmological equation
\begin{equation}
\ddot{a}(t) = -{{4\pi}\over3}G\left({\rho+3{p\over{c^2}}}\right) a(t)
\end{equation}
(where $a(t)$ is the scale factor at cosmic time $t$) it is easy to
derive that the considered region expands exponentially: $a(t)\propto
e^{t/t_f}$ with $t_f=(8\pi G\rho_f/3)^{-1/2}$; the epoch in which the
universe undergoes such exponential expansion is called {\it inflation}.

If $\rho_f$ is taken to be at about the grand unified theory scale, we
get $t_f\sim10^{-38}\;{\rm s}$, corresponding to an Hubble length of
$ct_f \sim 10^{-28}\;{\rm cm}$; if the inflationary phase is long enough
(a lower limit for this is about 65 e-folding times, corresponding to an
expansion by a factor $\sim10^{28}$), it smooths the metric, so that the
expanding region approaches a de Sitter flat space, regardless of
initial conditions. Furthermore, when inflation ends, the energy stored
in the inflaton\index{inflation} field is finally released, thermalizes and leads to the
hot mixture of particles assumed by the standard big bang picture.

\begin{figure}[t]
\epsfig{file=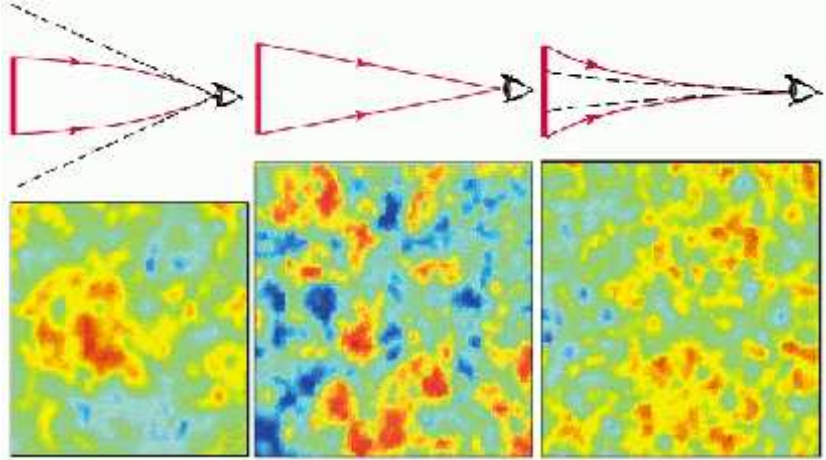,width=11truecm}
\caption{Effect of universe geometry on the observed angular size of
fluctuations in the CMBR\index{Cosmic Microwave Background}. If the universe is closed (left panel) ``hot
spots'' appear larger than actual size; if the universe is flat
(middle panel), ``hot spots'' appear with their actual size; if the
universe is open, ``hot spots'' appear smaller than their actual
size.}
\label{geometry}
\end{figure}

Inflation helps explaining several fundamental observations, which were
just assumed as initial conditions in non inflationary models:
\begin{enumerate}
\item{The Hubble expansion: repulsive gravity associated with false
vacuum is exactly the kind of force needed to set up a motion pattern in
which every two particles are moving apart with a velocity proportional
to their distance.}
\item{The homogeneity and isotropy: in ``classical'' cosmology the
remarkable uniformity of the Cosmic Microwave Background\index{Cosmic Microwave Background} Radiation (CMBR)
cannot be ``explained'', because different
regions of the present CMBR sky never were causally connected. Instead,
in inflationary models the whole CMBR sky was causally connected {\it before}
inflation\index{inflation} started, and uniformity can be established at that time.}
\item{The flatness problem: a flat Friedman-Robertson-Walker model
universe ({\it i.e.}\ with $\Omega(t)\equiv\rho_{\rm tot}(t)/\rho_{\rm
crit}(t)=1$, where $\rho_{\rm tot}(t)$ is the cosmological mean density,
including the ``dark energy'' term, and $\rho_{\rm crit}(t)=3H(t)^2/8\pi
G$, $H(t)\equiv \dot{a}(t)/a(t)$ being the Hubble parameter) always
remains flat, but if at an early time $\Omega$ was just slightly
different from $1$, the difference should have grown as fast as
$(\Omega-1) \propto t^{2/3}$. All the observational results ({\it e.g.}
the Bennet \etal 2003 WMAP result $\Omega_0=1.02 \pm 0.02$) show that at
present $\Omega$ is quite close to 1, implying that at the Planck time
($t\sim10^{-43}\;{\rm s}$) the closeness was really amazing.  Inflation
removes the need for this incredibly accurate fine tuning, since during
an inflationary phase $\Omega$ is driven towards 1 as $(\Omega-1)\propto
e^{-2Ht}$.}
\item{The absence of magnetic mono-poles: grand unified theories,
combined with classical cosmology, predict that the universe should be
largely dominated by magnetic mono-poles, which instead have never been
observed. Inflation\index{inflation} provides an explanation, since it can dilute the
mono-pole density to completely negligible levels.}
\item{the {\it anisotropy} properties of the CMBR\index{Cosmic Microwave Background} radiation: inflation
provides a prediction for the power spectrum of fluctuations, which
should be generated by quantum fluctuations and nearly scale-invariant,
a prediction in close agreement with the WMAP results (see the next
subsection).}
\end{enumerate}
 


A peculiar property of inflation\index{inflation} is that most inflationary models
predict that inflation does not stop everywhere at the same time, but
just in localized ``patches'' in a succession which continues eternally;
since each ``patch'' (such as the one we would be living in) can be
considered a whole universe, it can be said that inflation produces an
infinite number of universes.



\subsubsection{Primordial fluctuation evolution - Dark Matter\index{dark matter}}
Inflation\index{inflation} predicts the statistical properties of the density
perturbation field, defined as
\begin{equation}
\delta({\bf x}) \equiv {{\rho({\bf r}) - \bar\rho}\over{\bar\rho}}
\end{equation}
where ${\bf r}$ is the proper coordinate, ${\bf x}={\bf r}/a$ is the
comoving coordinate, and $\bar\rho$ is the mean matter density.

In fact, if we look at the Fourier components in terms of the comoving
wave-vectors ${\bf k}$ 
\begin{equation}
\delta({\bf k}) = \int{\delta({\bf x})\ e^{-i{\bf k}{\bf x}}\ d^3{\bf x}}
\end{equation}
the inflationary prediction is that the perturbation field is a
Gaussian random field, that the various ${\bf k}$ modes are independent,
and that the power spectrum $P(k)$ (where $k\equiv|{\bf k}|$) is close
to scale-invariant, {\it i.e.}, it is given by a power law
\begin{equation}
P(k) \equiv <|\delta_{\bf k}|^2> \propto k^{n_s},\qquad{\rm with}\ n_s\simeq 1.
\end{equation}

\begin{figure}[t]
\epsfig{file=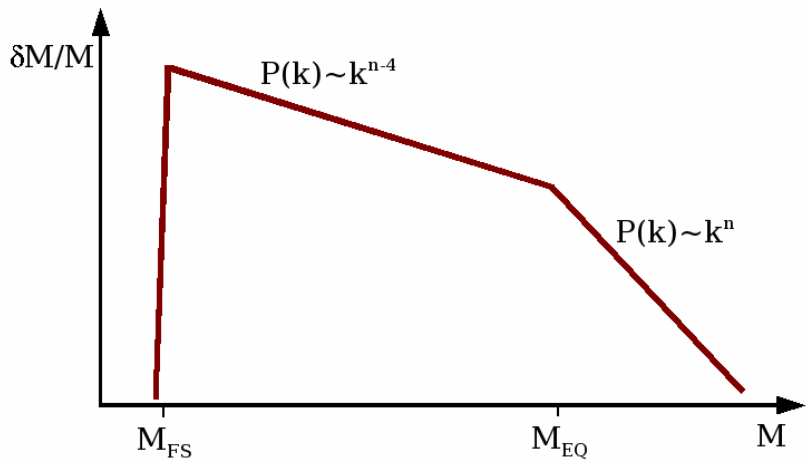,width=11truecm}
\caption{Schematic shape of the DM power spectrum after accounting
for the effects of free streaming\index{fluctuations!free streaming scale} and of the processing that takes
place for fluctuations entering the horizon before $t_{\rm eq}$.}
\label{powerspectrum}
\end{figure}

This prediction applies to the {\it primordial} power spectrum; in order
to make comparisons with observations, we need to include effects which
subsequently modified its shape:
\begin{enumerate}
\item{Free streaming: the dark matter\index{dark matter} particles are in motion, and since
they are believed to interact only through gravity, they freely
propagate from overdense to underdense regions, wiping out the density
perturbations. However, at any given cosmic time $t$, this will affect
only wavelengths smaller than the {\it free streaming}\index{fluctuations!free streaming scale} length, \ie\ the
proper distance $l_{\rm FS}(t)$ which a dark matter particle can have
travelled in time $t$. It can be shown (see {\it e.g.}\ Padmanabhan 1993,
\S4.6) that this scale depends primarily on the epoch $t_{\rm nr}$ when
the dark matter\index{dark matter} particles become non-relativistic: before that epoch
they move at the speed of light, and can cover a proper distance $l_{\rm
FS}(t_{\rm nr})\sim 2ct_{\rm nr}$, corresponding to a comoving distance
$\lambda_{\rm FS} = (a_0/a_{\rm nr})2ct_{\rm nr}$; after becoming
non-relativistic, particle motions become much slower than cosmological
expansion, and the evolution after $t_{\rm nr}$ only increases
$\lambda_{\rm FS}$ by a relatively small factor.  In turn, $t_{\rm nr}$
depends primarily on the mass $m_{\rm DM}$ of dark matter particles, so
that
\begin{equation}
\lambda_{\rm FS}\sim 5\times10^{-3}\;(\Omega_{\rm DM} h^2)^{1/3}
       \left({{m_{\rm DM}}\over{\rm 1\;GeV}}\right)^{-4/3}\ {\rm pc}
\end{equation}
corresponding to a mass scale
\begin{equation}
M_{\rm FS}\sim 6\times 10^{-15}\;(\Omega_{\rm DM} h^2)
       \left({{m_{\rm DM}}\over{\rm 1\;GeV}}\right)^{-4}\ \msunabel.
\end{equation}
Since the most favoured candidates for the DM particles are Weakly
Interacting Massive Particles (WIMPs) with mass between $0.1$ and
$100\;{\rm GeV}$, we probably have $M_{\rm FS}\lsim10^{-8} \msunabel$. Some
super-symmetric theories (\eg\ Schwartz \etal 2001, and more recently
Green, Hofmann \& Schwarz 2004) instead point towards $M_{\rm FS}\sim
10^{-6}\msunabel$, and Diemand, Moore \& Stadel (2005) used this result in
order to argue that the first structure which formed in the early
Universe were Earth-mass dark-matter\index{dark matter} haloes (but see also Zhao \etal
2005 for criticism about this result);}
\item{Growth rate changes: in general, perturbations grow
because of gravity; however, the details of the process change with
time, leaving an imprint on the final processed spectrum. The time
$t_{\rm eq}$ when the matter density becomes larger than the radiation
density is particularly important: before $t_{\rm eq}$, perturbations
with $\lambda$ larger than the Hubble radius $c/H(t)$ grow as
$\delta\propto a^2$, while the growth of smaller perturbations is almost
completely suppressed; after $t_{\rm eq}$ both large and small
fluctuations grow as $\delta\propto a$.  Because of these differences,
the size of the Hubble radius at $t_{\rm eq}$, $\lambda_{\rm eq}\simeq
13\;(\Omega h^2)^{-1} {\rm Mpc}$ (in terms of mass, $M_{\rm eq}\simeq
3.2\times10^{14} (\Omega h^2)^{-2}\;\msunabel$) separates two different
spectral regimes. In the wavelength range $\lambda_{\rm FS}\leq
\lambda\leq \lambda_{\rm eq}$ the growth of fluctuations pauses between
the time they enter the Hubble radius and $t_{\rm eq}$. As a result
the slope of the processed spectrum is changed, and $P(k)\propto k^{n_s-4}\simeq
k^{-3}$. Instead, at scales $\lambda>\lambda_{\rm eq}$ all the
fluctuations keep growing at the same rate at all times, and the shape
of the power spectrum remains similar to the primordial one, $P(k)\propto
k^{n_s}\simeq k^1$.}
\end{enumerate}

WMAP (Bennett \etal 2003) measured the spectral index,
obtaining $n_s=0.99\pm0.04$, and did not detect deviations from
gaussianity, both results in agreement with inflationary
predictions.

This kind of spectrum, in which fluctuations are typically larger on
small scales, leads naturally to hierarchical structure formation, since
small-scale fluctuations are the first to become non-linear\index{fluctuations!non-linear evolution} (\ie to
reach $\delta\sim1$), collapse and form some kind of astronomical
object.  It is also worth remarking that the very first objects, coming
from the highest peaks of $\delta({\bf x})$, are typically located where
modes $\delta({\rm k})$ of different wavelength make some kind of
``constructive interference'': the very first objects are likely to be
on top of larger ones, and they are likely to be clustered together,
rather than uniformly distributed. For this reason, it is also very
likely that the halos where these objects formed have long since been
incorporated inside larger objects, such as the bulges of $M_*$ galaxies
or the cD galaxy at the centre of galaxy clusters (see {\it e.g.}\ White
\& Springel 1999).

\begin{figure}[t]
\epsfig{file=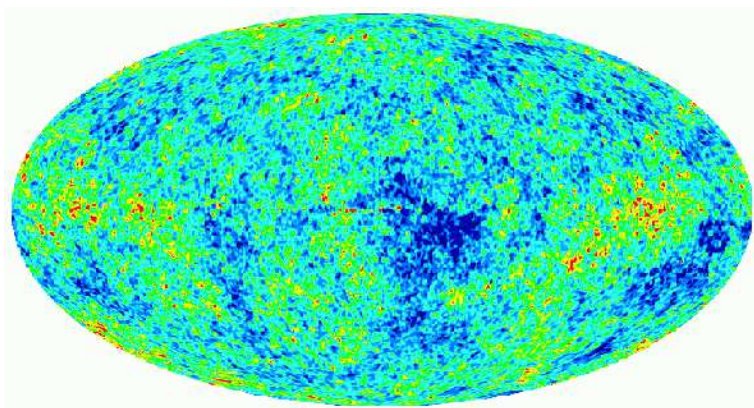,angle=0,width=10truecm,clip=}
\caption{Map of the temperature anisotropies in the CMB\index{Cosmic Microwave Background} as obtained by
combining the five bands of the WMAP satellite in order to minimise
foreground contamination (from Bennett \etal 2003).}
\label{wmap_map}
\end{figure}

\begin{figure}[t]
\epsfig{file=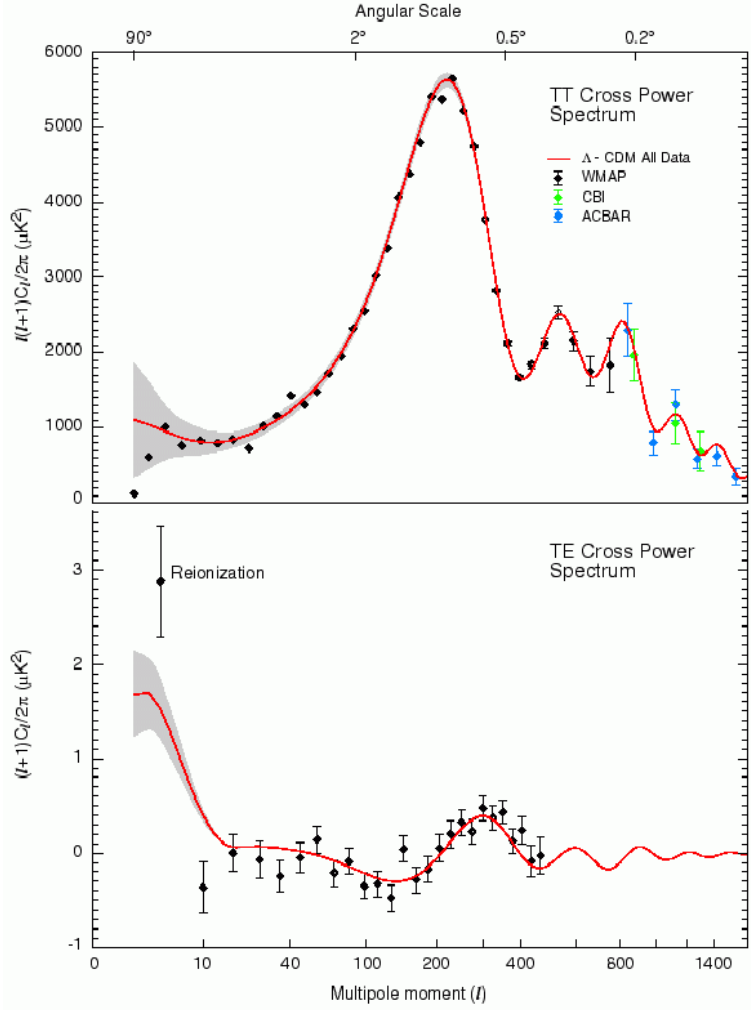,width=10truecm,clip=}
\caption{Angular CMB\index{Cosmic Microwave Background} power spectrum of temperature (top panel) and
temperature-polarization (bottom panel) as obtained by the WMAP
satellite (from Bennett \etal 2003). The line shows the best-fit with
a $\Lambda$CDM model, and grey areas represent cosmic variance.}
\label{wmap_angspec}
\end{figure}


\subsubsection{Fluctuation evolution - Baryons}

Before the equivalence epoch $t_{\rm eq}$ the baryons evolve in the same
way as dark matter\index{dark matter}. Instead, in the matter dominated era they behave
differently: we mentioned that all the dark matter fluctuations which
were not erased by free streaming\index{fluctuations!free streaming scale} grow as $\delta\propto a \propto
(1+z)^{-1}$, but this does not apply to baryons. In fact, baryons
decouple from radiation only at $t_{\rm dec}$, significantly later than
$t_{\rm eq}$ (we remind that $1+z_{\rm eq}\sim 10^4$, while $1+z_{\rm dec}
\simeq10^3$). The persistence of the coupling with radiation prevents
the growth of baryonic fluctuations on all scales; even worse, on
relatively small scales all the fluctuations in the baryonic component
are erased through a mechanism similar to free streaming. Such effect
takes place below the so-called {\it Silk scale}\index{fluctuations!silk scale} (Silk 1968), which is
given by the average distance that the photons (and the baryons coupled
with them) can diffuse before $t=t_{\rm dec}$; this translates into a
comoving distance
\begin{equation}
  \lambda_{\rm S} \simeq 3.5 \left({\Omega\over{\Omega_{\rm
  b}}}\right)^{1/2} (\Omega h^2)^{-3/4}\ {\rm Mpc}
\end{equation}
(where $\Omega_{\rm b}$ is the baryonic contribution to $\Omega$)
and encloses a mass
\begin{equation}
  M_{\rm S} \simeq 6.2\times10^{12} \left({\Omega\over{\Omega_{\rm
  b}}}\right)^{3/2} (\Omega h^2)^{-5/4}\ \msunabel.
\end{equation}

This result was a major problem for cosmology before the existence of
dark matter\index{dark matter} started to be assumed: it implies that in a purely baryonic
universe there should be no structures on a scale smaller than that of
galaxy clusters (if $\Omega=\Omega_{\rm b}\simeq 0.1$).  Furthermore,
even fluctuations which were not erased can grow only by a factor
$1+z_{\rm dec}$ between decoupling and present time, and this is not
enough to take a typical CMBR\index{Cosmic Microwave Background} fluctuation (of amplitude
$\delta\sim10^{-5}$) into the non-linear\index{fluctuations!non-linear evolution} regime.  The introduction of
Cold Dark matter solved this problem, since after recombination\index{Hydrogen!recombination} the
baryons are finally free to fall inside dark matter\index{dark matter} potential wells,
whose growth was unaffected by the photon drag. 

It can be found that after decoupling from radiation, the baryonic
fluctuations quickly ``reach'' the levels of dark matter fluctuations,
evolving as
\begin{equation}
\delta_{\rm b} = \delta_{\rm DM}\left({1-{{1+z}\over
    {1+z_{\rm dec}}}}\right),
\end{equation}
so that the existing dark matter\index{dark matter} potential wells ``regenerate'' baryonic
fluctuations, including the ones below the Silk scale\index{fluctuations!silk scale}.

This result is correct as long as pressure does not play a role, that is
for objects with a mass larger than the cosmological Jeans mass\index{Jeans mass} $M_{\rm
J}\propto T^{3/2}{\rho}^{-1/2}$. Such mass behaves differently at high
and low redshift. Before a redshift $z_{\rm Compton}\simeq 130$ we have
that the temperature of the baryons is still coupled to that of the CMB\index{Cosmic Microwave Background}
because of Compton scattering of radiation on the residual free
electrons; for this reason, $T_{\rm b}(z)\simeq T_{\rm CMB}(z)\propto (1+z)$,
and as $\rho(z)\propto (1+z)^3$ the value of $M_{\rm J}$ is constant:
\begin{equation}
M_{\rm J}(z)\simeq 1.35\times 10^5 \left({{\Omega_{\rm m} h^2}\over
  {0.15}}\right)^{-1/2}\;\msunabel \qquad
  ({\rm for}\ z_{\rm dec}\gsim z\gsim z_{\rm Compton})
\end{equation}
where $\Omega_{\rm m}=\Omega_{\rm b}+\Omega_{\rm DM}$ is the total
matter density (baryons plus dark matter\index{dark matter}).
At lower redshifts the baryon temperature is no longer locked to that of
the CMB\index{Cosmic Microwave Background} and drops adiabatically as $T_{\rm b}\propto (1+z)^2$. At such
redshifts the Jeans mass\index{Jeans mass} evolves as
\begin{eqnarray}
M_{\rm J}(z)\simeq 5.7\times 10^3
   \left({{\Omega_{\rm m} h^2}\over {0.15}}\right)^{-1/2} \;
   \left({{\Omega_{\rm b}h^2}\over{0.022}}\right)^{-3/5} \;
   \left({{1+z}\over{10}}\right)^{3/2}\;\msunabel \qquad
   \\({\rm for}\ z\lsim z_{\rm Compton}).\nonumber 
\end{eqnarray}

\begin{figure}[t]
\epsfig{file=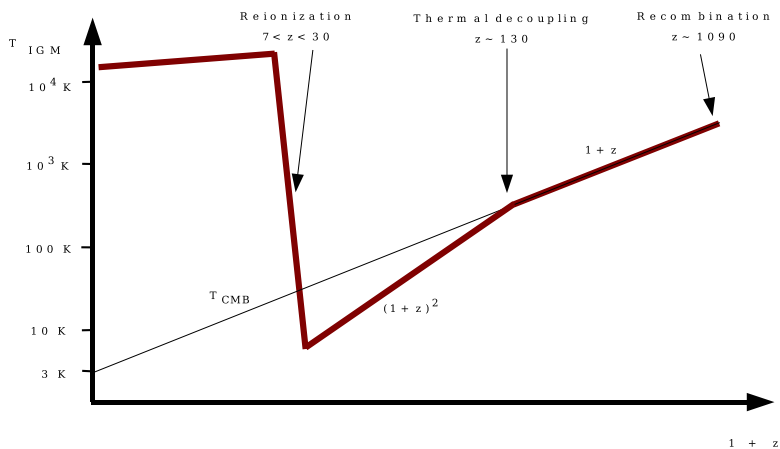,width=11truecm,clip=}
\caption{Schematic evolution of the temperature of the Intergalactic Medium
(IGM) as a function of redshift: after recombination and before
thermal decoupling (at $z_{\rm Compton}$), $T_{\rm IGM}$ is locked to
$T_{\rm CMBR}$ and evolves as $(1+z)$. After thermal decoupling
$T_{\rm IGM}$ is no more locked to the radiation temperature, and
drops adiabatically as $(1+z)^2$, until the first non-linear objects
collapse, and emit light which is able to re-ionize the universe,
raising $T_{\rm IGM}$ above $\sim 10^4\;{\rm K}$.}
\label{thermal_hist}
\end{figure}


\subsection{From fluctuations to cosmological structures}
\subsubsection{Non-linear evolution\index{fluctuations!non-linear evolution}}
When the density contrast $\delta$ becomes comparable to unity, the
evolution becomes non-linear, and the Fourier modes no more evolve
independently. The easiest way to study this phase is through the ``real
space'' $\delta({\bf x})$ (rather than its Fourier components
$\delta({\bf k})$), considering the idealized case of the collapse of a
spherically symmetric overdensity, and in particular the collapse of a
{\it top-hat} fluctuation. It is well known that, through some simple
further assumption (such as that ``spherical shells do not cross''), the
time evolution of the radius $R$ of a top-hat perturbation (see \eg
Padmabhan 1993, \S 8.2 for the treatment of a slightly more general
case) can be written down as
\begin{equation}
R(t)={R_i\over2}{{1+\delta_i}\over{\delta_i-(\Omega_i^{-1}-1)}}[1-\cos\theta(t)]
\label{tophatradius}
\end{equation}
where $R_i=R(t_i)$, $\delta_i=\delta(t_i)$ and $\Omega_i=\Omega(t_i)$
are the ``initial conditions'' for the fluctuation evolution, and
$\theta$ is defined by the equation
\begin{equation}
[\theta(t)-\sin\theta(t)]{{1+\delta_i}\over
{2H_i\Omega_i^{1/2}[\delta_i-(\Omega_i^{-1}-1)]^{3/2}}} \simeq  t
\end{equation}
where again $H_i$ is the Hubble parameter at $t_i$, and the last
approximate equality is valid only as long as $\delta_i\ll1$ (that is, a
sufficiently early $t_i$ must be chosen).  The fluctuation radius $R$
reaches a maximum $R_{\rm ta}$ at the so-called {\it turn-around} epoch
(when $\theta=\pi$) when the overdense region finally detaches itself
from the Hubble flow and starts contracting. However, while
eq. (\ref{tophatradius}) suggests an infinite contraction to $R=0$ (when
$\theta=2\pi$), the violent relaxation process (Lynden-Bell 1967)
prevents this from happening, leading to a configuration in virial
equilibrium at $R_{\rm vir}\simeq R_{\rm ta}/2$.

Here, we summarise some well known, useful findings of this model.

First of all, combining the evolution of the background density
evolution and of eq. (\ref{tophatradius}) it is possible to
estimate the density contrast evolution
\begin{equation}
\delta(t)={9\over2}{{(\theta-\sin\theta)^2}\over{(1-\cos\theta)^3}}-1
\end{equation}
which leads to some noteworthy observation, such as that the density
contrast at turn-around is $\delta_{\rm ta}=(9\pi^2/16)-1\simeq4.6$,
which at virialization becomes $\delta_{\rm vir}=\Delta_{\rm c}$,
where it can be usually assumed that $\Delta_{\rm c}\simeq18\pi^2$
but sometimes higher order approximations are necessary, such as the
one in Bryan \& Norman 1998,
\begin{equation}
\Delta_{\rm c}=18\pi^2 + 82(1-\Omega_m^z) -39(1-\Omega_m^z)^2
\end{equation}
with
\begin{equation}
\Omega_m^z={{\Omega_{\rm m} (1+z)^3}\over
{\Omega_{\rm m} (1+z)^3+\Omega_\Lambda+\Omega_k(1+z)^2}}
\end{equation}
where $\Omega_\Lambda$ is the dark energy density, and $\Omega_k =
1-\Omega_{\rm m}-\Omega_\Lambda$ is the curvature.

From this, it is possible to estimate the virial radius
\begin{equation}
R_{\rm vir} \simeq 0.784 \left({M\over{10^8 h^{-1} \msunabel}}\right)^{1/3}
\left({{\Omega_{\rm m}\over\Omega_m^z}{\Delta_c\over{18\pi^2}}}\right)^{-1/3}
\left({{1+z}\over{10}}\right)^{-1} h^{-1}\;{\rm kpc},
\end{equation}
the circular velocity for such an halo
\begin{equation}
V_{\rm circ} \simeq 23.4 \left({M\over{10^8 h^{-1} \msunabel}}\right)^{1/3}
\left({{\Omega_{\rm m}\over\Omega_m^z}{\Delta_c\over{18\pi^2}}}\right)^{1/6}
\left({{1+z}\over{10}}\right)^{1/2}\;{\rm km\,s^{-1}},
\end{equation}
and the virial temperature
\begin{equation}
T_{\rm vir} \simeq 19800 \left({M\over{10^8 h^{-1} \msunabel}}\right)^{2/3}
\left({{\Omega_{\rm m}\over\Omega_m^z}{\Delta_c\over{18\pi^2}}}\right)^{1/3}
\left({{1+z}\over{10}}\right) \left({\mu\over{0.6}}\right) \;{\rm K}.
\end{equation}

\subsubsection{The Press-Schechter formalism\index{Press-Schechter formalism}} 
\begin{figure}[t]
\epsfig{file=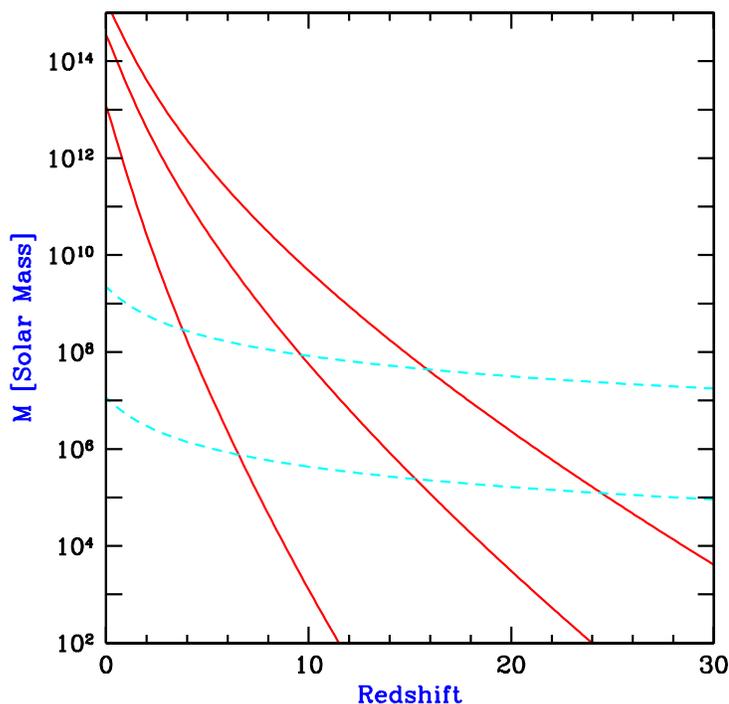,width=10truecm,clip=}
\caption{Characteristic mass of $1\sigma$ (bottom solid curve),
$2\sigma$ (middle solid curve) and $3\sigma$ (top solid curve)
collapsing halos as a function of redshift. These were obtained from
the Eisenstein \& Hu (1999) power spectrum, assuming an
$\Omega_\Lambda=0.7$, $\Omega_{\rm m}=0.3$ cosmology. The dashed curves show
the minimum mass\index{minimum mass} which is required for the baryons to be able to cool
and collapse (see next section) in case of pure atomic cooling (upper
curve) and of molecular cooling\index{cooling} (lower curve).
(from Barkana \& Loeb 2001).}
\label{press_schechter_mass}
\end{figure}

\begin{figure}[t]
\epsfig{file=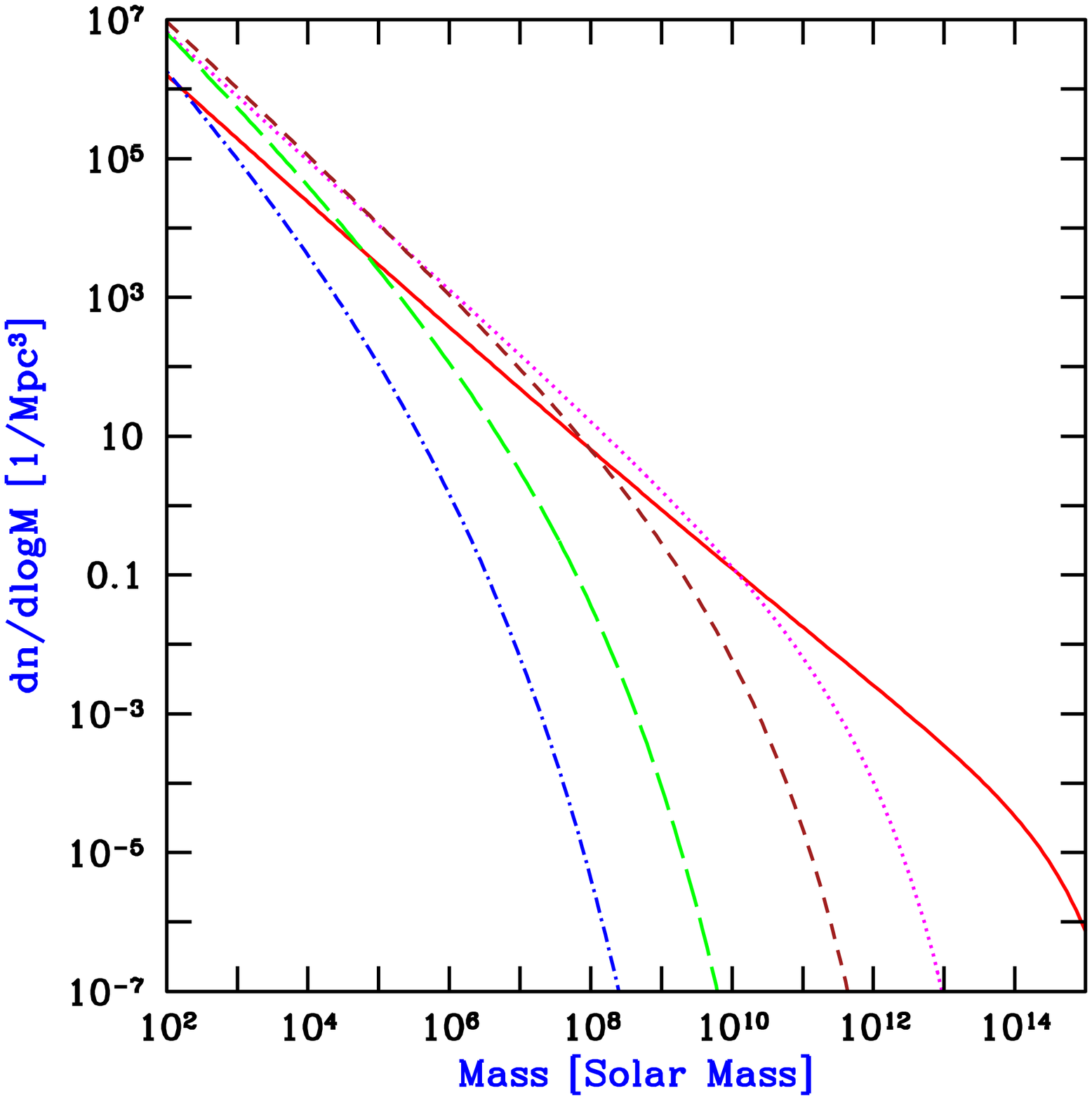,width=10truecm,clip=}
\caption{Halo mass functions at several redshifts (from bottom left to
  bottom right, $z=30$, $z=20$, $z=10$, $z=5$ and $z=0$,
  respectively). The assumed power spectrum and cosmology are the same as in fig. \ref{press_schechter_mass}. (from Barkana \& Loeb 2001).}
\label{press_schechter_abundance}
\end{figure}

The simple top-hat model is at the core of the so-called Press-Schechter
formalism\index{Press-Schechter formalism} (Press \& Schechter 1974, but see also the contribution by
Sommerville in this same book), which predicts the density of
virialized halos of a given mass at a given redshift. This model assumes
that the distribution of the smoothed density field $\delta_M$ (where
$M$ is the mass scale of the smoothing) at a certain redshift $z_0$ is
Gaussian with a variance $\sigma_M$, so that the probability of having
$\delta_M$ larger than a given $\delta_{\rm crit}$ is
\begin{equation}
P(\delta_M>\delta_{\rm crit}) = \int_{\delta_{\rm crit}}^\infty {
{1\over{(2\pi)^{1/2}\sigma_M}} e^{-{x^2\over{2\sigma_M^2}}} dx};
\end{equation}
a common choice is $z_0=0$, requiring $\delta_M$ to be estimated through
a purely linear evolution\index{fluctuations!linear evolution} of the primordial power spectrum.

The Press-Schechter\index{Press-Schechter formalism} model then chooses a $\delta_{\rm crit}=\delta_{\rm
crit}(z)$ (but it is also possible to assume a constant $\delta_{\rm
crit}$ and make $\sigma_M$ a function of redshift; see \eg\ Viana \&
Liddle 1996) and assumes that this probability (multiplied by a factor
of 2 - see Bond \etal 1991 for an explanation of this extra factor) also
gives the fraction mass which at a redshift $z$ is inside virialized
halos of mass M or larger.  This can be differentiated over $M$ in order
to get the mass distribution at each given redshift
\begin{equation}
{{dn}\over{dM}}={2\over{(2\pi)^{1/2}}} {\rho_m\over M}
{{d ln(1/\sigma_M)}\over{dM}} {{\delta_{\rm crit}(z)}\over\sigma_M}
e^{-{{\delta_{\rm crit}(z)^2}\over{2\sigma_M^2}}}
\end{equation}

In this way, the abundance of halos is completely determined through the
two functions $\delta_{\rm crit}(z)$ and $\sigma_M$.  The first one is
commonly written as $\delta_{\rm crit}(z)=\delta_0/D(z)$, where $D(z)$
is the growth factor ($D(z)\simeq(1+z)^{-1}$ for Einstein-de Sitter
models; see Peebles 1993 for a more general expression), coming from
cosmology; instead $\delta_0$ is usually taken to be $1.686$, since the
top-hat model predicts that an object virializes at the time when the
linear theory estimates its overdensity at $\delta=1.686$.  Instead
$\sigma_M$ depends from the power spectrum; for example, figures
\ref{press_schechter_mass} and \ref{press_schechter_abundance} are based
on the Eisenstein \& Hu (1999) results\footnote{The authors of this
paper also provide some very useful codes for dealing with the power
spectrum and the Press-Schechter formalism\index{Press-Schechter formalism} at the web page
http://background.uchicago.edu/$\sim$whu/transfer/transferpage.html.}.


\section{Primordial gas properties}

\subsection{Cooling}
The typical densities reached by the gas after virialization (of the
order of $n_{\rm B}\equiv \rho_{\rm B}/m_{\rm H} \sim 0.01 \Omega_{\rm
b} [(1+z_{\rm vir})/10]^3 {\rm cm^{-3}}$ are far too low for the gas to
condense and form an object like a star. The only way to proceed further
in the collapse and in the formation of luminous objects is to remove
the gas thermal energy through radiative cooling\index{cooling}.

For this reason, cooling processes are important in determining where,
when and how the first luminous objects will form.

\begin{figure}[t]
\epsfig{file=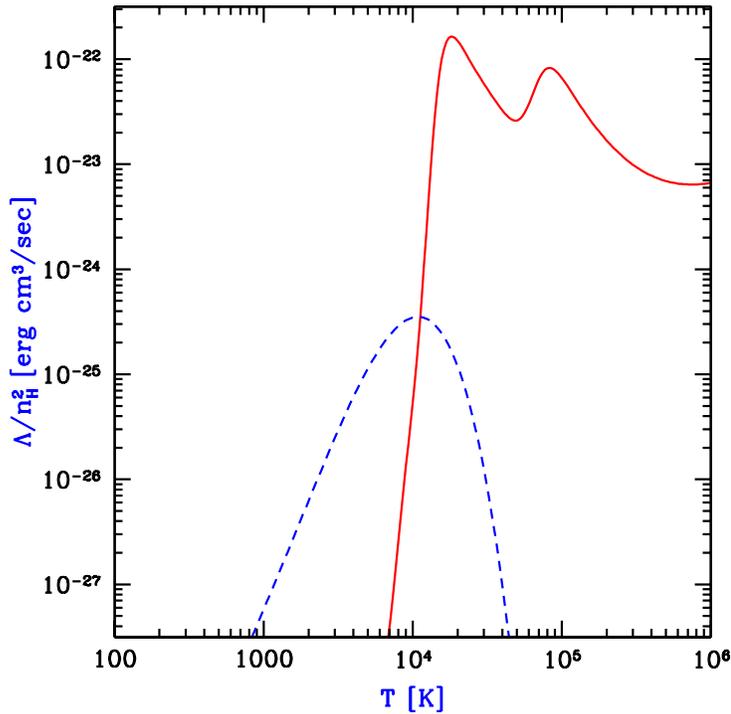,width=10truecm,clip=}
\caption{Cooling rate per atomic mass unit of\index{cooling!metal-free gas} metal-free gas, as a
function of temperature. The solid line shows assumes the gas to be
completely atomic (the two peaks correspond to H and He excitations,
while the high temperature tail is dominated by free-free processes) and
drops to about zero below $T\sim10^4\;{\rm K}$; the dashed line shows
the contribution of a small ($f_{\rm H_2} = 10^{-3}$) fraction of
molecular hydrogen\index{Hydrogen}, which contributes extra-cooling in the range
$100\;{\rm K}\lsim T \lsim 10^4\;{\rm K}$ (from Barkana \& Loeb 2001).}
\label{metalfree_cooling}
\end{figure}

In Fig. \ref{metalfree_cooling} it is possible to see that the cooling\index{cooling!metal-free gas}
of primordial (\ie metal-free) {\it atomic} gas at temperatures below
$\sim10^4\;{\rm K}$ is dramatically inefficient, because in that
temperature range the gap between the fundamental and the lowest excited
levels ($\simeq 10.2\;{\rm eV}$ for H atoms) is so much larger than the
thermal energy $\sim k_{\rm B}T\lsim 1\;{\rm eV}$ that very few atoms
get collisionally excited, and very few photons are emitted through the
corresponding de-excitations.

This is important because in all hierarchical scenarios the first
objects to virialize are the smallest ones, and such halos have
the lowest virial temperatures. If the primordial gas were completely
atomic, the first luminous objects would probably form relatively late
($z\lsim10-15$), in moderately massive halos ($M\sim10^8\;\msunabel$) with
$T_{\rm vir}\gsim 10^4\;{\rm K}$.

However, it is also possible to see from Fig. \ref{metalfree_cooling} that
the presence of molecules in small amounts ($f_{\rm H_2}\equiv 2n_{\rm
H_2}/(n_{\rm H}+2n_{\rm H_2})\gsim 5\times10^{-4}$; the dashed curve in
Fig. \ref{metalfree_cooling} was obtained assuming $f_{\rm
H_2}=10^{-3}$) can dramatically affect the cooling\index{cooling} properties of
primordial gas at low temperatures, making low mass halos virializing at
high redshift ($z\gsim 20$) the most likely sites for the formation of
the first luminous objects. 

\subsection{Molecular cooling\index{cooling}}

In the current scenario for the formation of primordial objects, the
most relevant molecule is H$_2$.  The only reason for this is the high
abundance of \HH\index{HH} when compared with all other molecules. In fact, the
radiating properties of an \HH molecule are very poor: because of the
absence of a dipole moment, radiation is emitted only through weak
quadrupole transitions. In addition, the energy difference between the
\HH ground state and the lowest \HH excited roto-vibrational levels is
relatively large ($\Delta E_{01}/k_{\rm B}\gsim 200\;{\rm K}$, between
the fundamental and the first excited level; however, such transition is
prohibited by quadrupole selection rules, and the lowest energy gap for
a quadrupole transition is $\Delta E_{02}/k_{\rm B}\simeq 510\;{\rm
K}$), further reducing the cooling efficiency at low temperatures.

\begin{figure}[t]
\epsfig{file=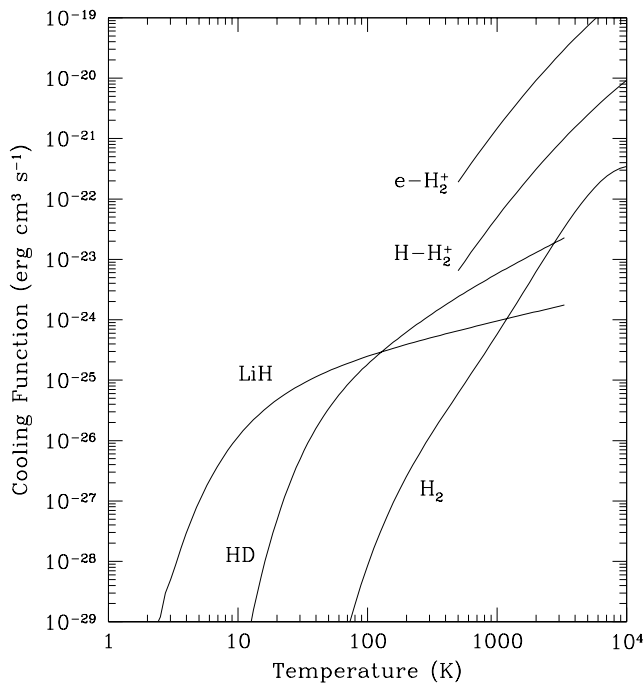,width=10truecm,clip=}
\caption{Comparison of the low density ($n\lsim 10\;{\rm cm^{-3}}$)
cooling rates per molecule of several molecular species, in particular
\HH, HD and LiH (from GP98). Note that at $T\sim100\;{\rm
K}$ both HD and LiH molecules are more than $10^3$ times more efficient
coolants than \HH molecules, but this difference is believed to be
compensated by the much higher \HH abundance (see {\it e.g.}, Bromm
\etal 2002). The plot also shows the cooling due to H-\HHP and
$e^-$-\HHP collisions, but these contributions are never important
because of the very low \HHP abundance.}
\label{species_cooling}
\end{figure}

Apart from \HH the most relevant molecular coolants are HD and LiH. In
the following, we will briefly list the cooling\index{cooling} rates (mainly taken from
Galli \& Palla 1998, hereafter GP98; see also Hollenbach \& McKee 1979,
Lepp \& Shull 1983, 1984 and Martin \etal 1996, Le Bourlot \etal 1999,
Flower \etal 2000) of \HH and of the other two possibly relevant
species\footnote{In Fig. \ref{species_cooling} the cooling rate for \HHP
is shown, too. But while it is still marginally possible that HD or LiH
cooling\index{HD cooling}\index{LiH cooling} can play some kind of role in the primordial universe, this is
much more unlikely for \HHP, mainly because its under-abundance with
respect to \HH is always much larger than the difference in the cooling
rates; for this reasons we choose to omit a detailed discussion of \HHP
cooling.}\index{cooling}.

We note that Bromm \etal 1999 included HD cooling\index{HD cooling} in some of
their simulations but found that it never accounted for more than
$\sim10\%$ of \HH cooling; however, they did not completely rule out the
possibility that a $n_{\rm HD}/n_{\rm H_2}$ ratio substantially larger
than the equilibrium value (close to $n_{\rm D}/n_{\rm H}\sim 10^{-5}$)
could change this conclusion. Also, in present-day simulations of
primordial star formation the gas temperature never goes below a few
hundreds degrees: in case it did, even a tiny amount of LiH\index{LiH cooling} could be
enough to dominate the cooling\index{cooling} rate.

\begin{figure}[t]
\epsfig{file=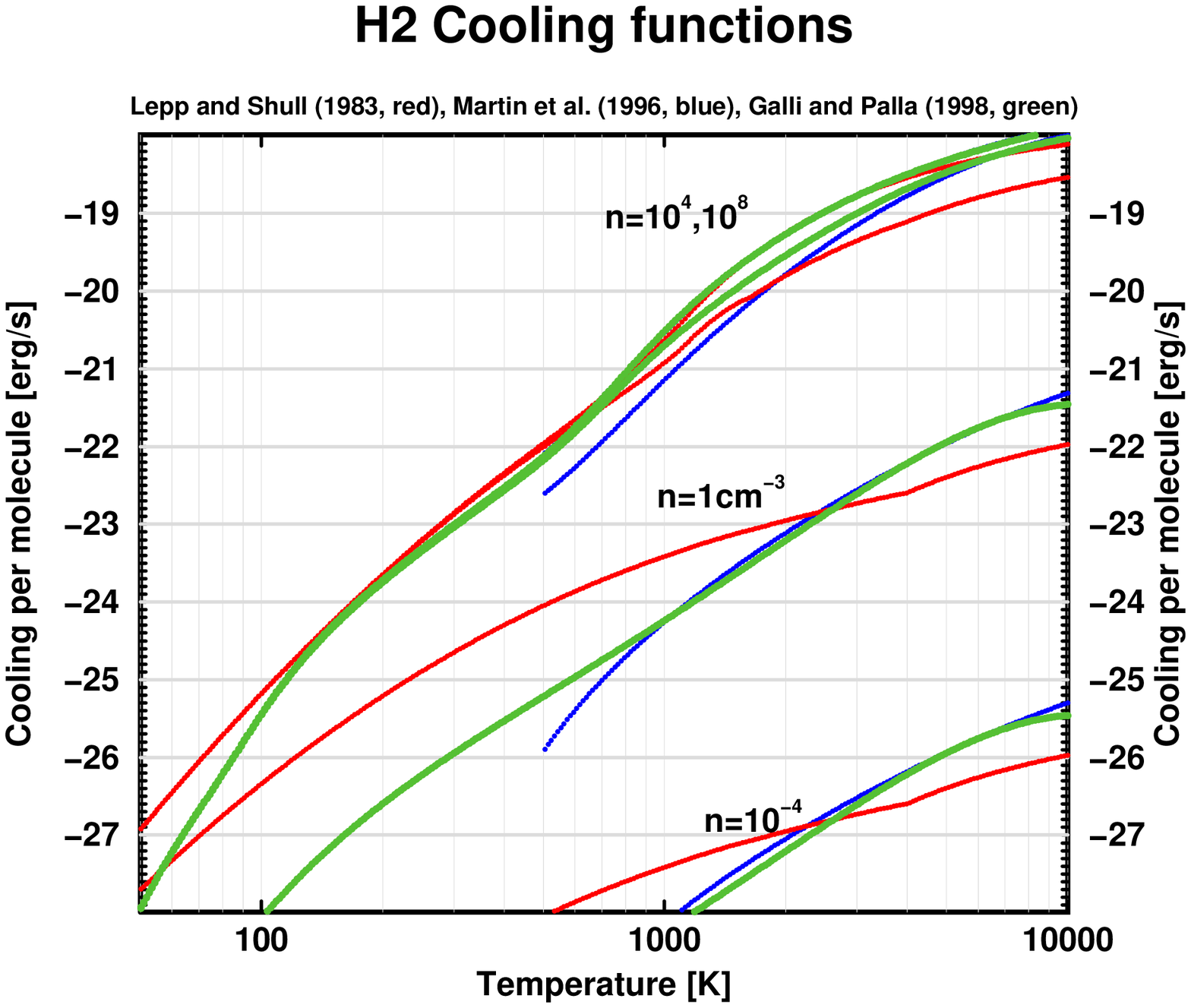,width=10truecm,clip=}
\caption{Cooling rate per \HH molecule as computed by different authors
(Lepp \& Shull 1983, Martin \etal 1996, GP98) as a
function of temperature and for different densities; note that for
$n\gsim 10^4\;{\rm cm^{-3}}$ the cooling rate is almost independent of
density.}
\label{H2_cool}
\end{figure}

\subsubsection{\HH cooling rate\index{cooling}}
The \HH cooling rate {\it per molecule} $\Lambda_{\rm H_2}(\rho,T)$
can be conveniently expressed in the form:
\begin{equation}
\Lambda_{\rm H_2}(\rho,T) = {{\Lambda_{\rm H_2,LTE}(T)}\over
{1+{{\Lambda_{\rm H_2,LTE}(T)}\over{n_{\rm H}\Lambda_{\rm
H_2,\rho\rightarrow0}(T)}}}}
\end{equation}
where $\Lambda_{\rm H_2,LTE}(T)$ and $n_{\rm H}\Lambda_{\rm
H_2,\rho\rightarrow0}$ are the high and low density limits of the
cooling rate (which apply at $n \gsim 10^4\; {\rm cm^{-3}}$ and
at $n \lsim 10^2\; {\rm cm^{-3}}$, respectively).

The high density (or LTE, from the Local Thermal Equilibrium assumption
which holds in these conditions) limit of the cooling rate per \HH
molecule is given by Hollenbach \& McKee (1979):
\begin{eqnarray}
\Lambda_{\rm H_2,LTE}(T)={{9.5\times10^{-22}T_3^{3.76}} \over
  {1+0.12\,T_3^{2.1}}} e^{-\left({0.13}\over{T_3}\right)^3} +
  3\times10^{-24} e^{-{{0.51}\over{T_3}}} + \nonumber \\
  + 6.7\times10^{-19} e^{-{{5.86}\over{T_3}}} +
  1.6\times10^{-18} e^{-{{11.7}\over{T_3}}}\;
  {\rm erg\,s^{-1}}
\end{eqnarray}
where $T_3 \equiv T/(1000\;{\rm K})$.  Note that the first row in the
formula accounts for rotational cooling\index{cooling}, while the second row accounts for
the first two vibrational terms.

For the low density limit, GP98 found that in the
relevant temperature range ($10\;{\rm K}\leq T \leq 10^4\;{\rm K}$) the
cooling rate $\Lambda_{\rm H_2,\rho\rightarrow0}$ is independent from
density, and is well approximated by
\begin{eqnarray}
\log \Lambda_{\rm H_2,\rho\rightarrow0}(T) \simeq 
  -103.0 + 97.59 \log T - 48.05 (\log T)^2 + \nonumber \\
  10.80 (\log T)^3 - 0.9032 (\log T)^4
\end{eqnarray}
where $T$ and $\Lambda_{\rm H_2,\rho\rightarrow0}$ are expressed in
$K$ and ${\rm erg\,s^{-1}\,cm^3}$, respectively.

Note that even if both $\Lambda_{\rm H_2,LTE}$ and
$\Lambda_{\rm H_2,\rho\rightarrow0}$ do not depend on density,
$\Lambda_{\rm H_2}(\rho,T)$ is independent of $\rho$ only in the high
density limit.

\subsubsection{HD\index{HD cooling} and LiH\index{LiH cooling} cooling rates}
The cooling rates of HD and of LiH are more complicated (see Flower
\etal 2000 for HD and Bogleux \& Galli 1997 for LiH), but in the low
density limit (and in the temperature range $10\;{\rm K}\leq T \leq
1000\;{\rm K}$) it is possible to use the relatively simple expressions
given by GP98.

For HD, we have that the low density limit of the cooling rate per
molecule, $\Lambda_{\rm HD,\rho\rightarrow0}$, is:
\begin{equation}
\Lambda_{\rm HD,\rho\rightarrow0}(T) \simeq 
2\gamma_{10}E_{10}e^{-{{E_{10}}\over{k_{\rm B}T}}} +
(5/3)\gamma_{21}E_{21}e^{-{{E_{21}}\over{k_{\rm B}T}}}
\end{equation}
where $E_{10}$ and $E_{21}$ are the energy gaps between HD levels 1 and
0 and levels 2 and 1, respectively; they are usually expressed as
$E_{10}=k_{\rm B} T_{10}$ and $E_{21}=k_{\rm B} T_{21}$, with
$T_{10}\simeq128\;{\rm K}$ and $T_{21}\simeq255\;{\rm K}$,

$\gamma_{10}$ and $\gamma_{21}$ are the approximate collisional
de-excitation rates for the 1-0 and 2-1 transitions, and are given by
\begin{eqnarray}
\gamma_{10} \simeq &
4.4\times10^{-12}+3.6\times10^{-13}
  T^{0.77}\\
\gamma_{21} \simeq &
4.1\times10^{-12}+2.1\times10^{-13}
  T^{0.92}
\end{eqnarray}
where we use the numerical value of T (in Kelvin) and the rates are
expressed in ${\rm cm^3\,s^{-1}}$.

For LiH instead we have that the same density limit of the cooling\index{cooling} rate
per molecule, $\Lambda_{\rm LiH,\rho\rightarrow0}$ can be fitted by:
\begin{eqnarray}
\log_{10}(\Lambda_{\rm LiH,\rho\rightarrow0}) = &
c_0 + c_1\log_{10} T + c_2(\log_{10} T)^2 + \nonumber\\
& c_3(\log_{10} T)^3 + c_4(\log_{10} T)^4
\end{eqnarray}
where $c_0=-31.47,\ c_1=8.817,\ c_2=-4.144,\ c_3=0.8292$ and
$c_4=-0.04996$, assuming that $T$ is expressed in Kelvin, and that
$\Lambda_{\rm LiH,\rho\rightarrow0}$ is expressed in
${\rm erg\,cm^3\,s^{-1}}$.

\subsubsection{Cooling at high densities: Collision Induced Emission\index{cooling!collision induced emission}}

\begin{figure}[t]
\epsfig{file=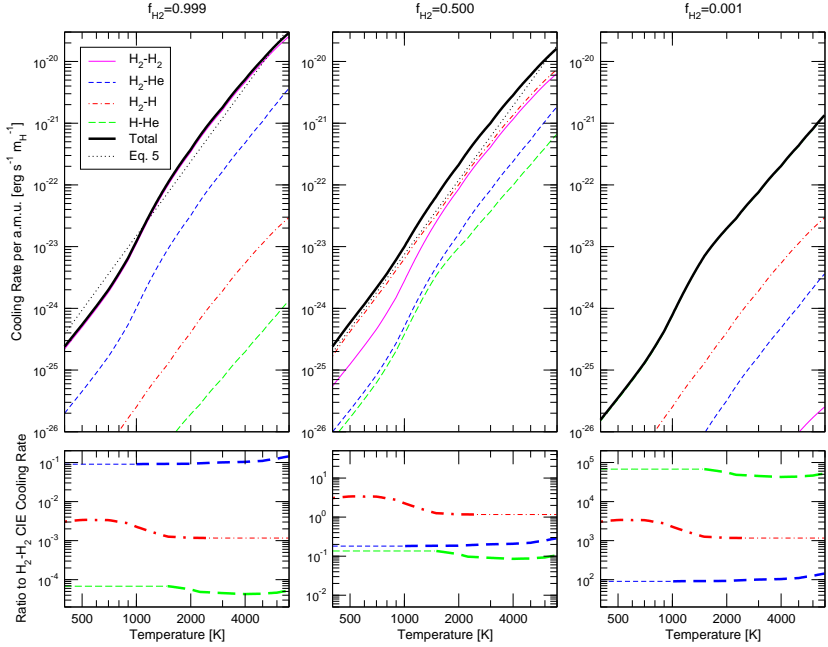,width=9truecm,angle=270}
\caption{CIE cooling for different collisions and different molecular
fractions(0.999, 0.5, 0.001). Top panels show the cooling rates per unit
mass: the thick solid line is the total CIE cooling, while thin lines
show the various components: \HH-\HH (solid), \HH-He (short dashed),
\HH-H (dot dashed) and H-He (long dashed); the dotted line shows the
results of the approximate formula given in the text. In the bottom
panels the ratios of the various component to \HH-\HH CIE is shown. All
quantities were calculated assuming $n=10^{14}\;{\rm cm^{-3}}$ and
$X=0.75$ (from Ripamonti \& Abel 2004).}
\label{cie_cooling}
\end{figure}

During the formation of a protostar an important role is played by the
so called Collision-Induced Emission (CIE; very often known as
Collision-Induced Absorption, or CIA), a process which requires pretty
high densities ($n=\rho/m_{\rm H}\gsim10^{13}-10^{14}\;{\rm cm^{-3}}$)
to become important (see Lenzuni \etal 1991, Frommold 1993, Ripamonti \&
Abel 2004). In fact, Collision-Induced Emission takes place when a
collision between two \HH molecules (or \HH and H, or \HH and He, or
even H and He) takes place: during the collision, the interacting pair
briefly acts as a ``super-molecule'' with a nonzero electric dipole, and
a much higher probability of emitting (CIE) or absorbing (CIA) a photon
than an unperturbed \HH molecule (whose dipole is 0). Because of the
very short durations of the collisions, ($\lsim10^{-12}\;{\rm s}$), such
mechanism can compete with ``normal'' emission only in high density
environments. Furthermore, because of the short durations of the
interactions, collision-induced lines become very broad and merge into a
continuum: the \HH CIE spectrum only shows broad peaks corresponding to
vibrational bands, rather than a large number of narrow roto-vibrational
lines. This is also important because self-absorption is much less
relevant than for line emission, and in primordial proto-stars
simulations CIE\index{cooling!collision induced emission} cooling can be treated with the optically thin
approximation up to $n\sim10^{16}\;{\rm cm^{-3}}$ (for \HH lines, the
optically thin approximation breaks at about $n\sim10^{10}\;{\rm
cm^{-3}}$).  In figure \ref{cie_cooling} we show the CIE cooling rate
for a gas with $n=10^{14}\;{\rm cm^{-3}}$, as a function of temperature
and for different chemical compositions; at temperatures between $400$
and $7000\;{\rm K}$.
For \HH abundances $f_{\rm H_2}\equiv 2n_{\rm H_2}/(n_{\rm H^+}+n_{\rm
H}+2n_{\rm H_2}) \gsim 0.5$ the total CIE\index{cooling!collision induced emission} cooling rate can be
approximated by the simple expression (see Ripamonti \& Abel 2004)
\begin{equation}
L_{\rm CIE}(\rho,T,X,f_{\rm H_2}) \simeq 0.072 \rho T^4 X f_{\rm H_2}
\;{\rm erg\,g^{-1}\,s^{-1}}
\end{equation}
where the density is in $\rm g\,cm^{-3}$, the temperature is in K, and
$X\simeq0.75$ is the hydrogen\index{Hydrogen} fraction (by mass).


\subsection{Chemistry\index{Chemistry}}

\begin{table}[b]
\caption{Reaction rates for some of the most important reactions in
primordial gas, plus the main reactions involved in the formation of
HD. In the formulae, $T_\gamma$ is the temperature of the radiation
field (for our purposes, the temperature of the CMB\index{Cosmic Microwave Background} radiation) and
temperatures need to be expressed in Kelvin; $j(\nu_{\rm LW})$ is the
radiation flux (in ${\rm erg\,s^{-1}\,cm^{-2}}$) at the central
frequency $\nu_{\rm LW}$ of the Lyman-Werner bands, $h\nu_{\rm
LW}=12.87\;{\rm eV}$.  The rates come from compilations given by Tegmark
\etal 1997 (reactions 1-7), Palla \etal 1983 (reactions 8-13), Abel
\etal 1997 (reaction 14) and Bromm \etal 2002 (reactions 15-19).}
\begin{tabular}{cccl}
\hline\hline
Reaction & & & Rate\\
\hline
${\rm H}^+ + e^-$ &  $\rightarrow$ & ${\rm H} + h\nu$ &
$k_1 \simeq 1.88\times10^{-10} T^{-0.644} \;{\rm cm^3\,s^{-1}}$\\

${\rm H} + e^-$ & $\rightarrow$ & ${\rm H}^- + h\nu$ &
$k_2 \simeq 1.83\times10^{-18} T^{0.88} \;{\rm cm^3\,s^{-1}}$\\

${\rm H}^- + {\rm H}$ & $\rightarrow$ & ${\rm H}_2 + e^- $ &
$k_3 \simeq 1.3\times10^{-9} \;{\rm cm^3\,s^{-1}}$\\

${\rm H}^- + h\nu$ & $\rightarrow$ & ${\rm H} + e^-$ &
$k_4 \simeq 0.114 T^{2.13} e^{-8650/T_\gamma} \;{\rm cm^3\,s^{-1}}$\\

${\rm H}^+ + {\rm H}$ & $\rightarrow$ & ${\rm H}_2^+ + h\nu$ &
$k_5 \simeq 1.85\times10^{-23} T^{1.8} \;{\rm cm^3\,s^{-1}}$\\

${\rm H}_2^+ + {\rm H}$ & $\rightarrow$ & ${\rm H}_2- + {\rm H}^+$ &
$k_6 \simeq 6.4\times10^{-10} \;{\rm cm^3\,s^{-1}}$\\

${\rm H}_2^+ + h\nu$ & $\rightarrow$ & ${\rm H}^+ + {\rm H}$ &
$k_7 \simeq 6.36\times10^5 e^{-71600/T_\gamma} \;{\rm cm^3\,s^{-1}}$\\

${\rm H} + {\rm H} + {\rm H}$ & $\rightarrow$ & ${\rm H}_2 + {\rm H}$ &
$k_8 \simeq 5.5\times10^{-29} T^{-1} \;{\rm cm^6\,s^{-1}}$\\

${\rm H}_2 + {\rm H}$ & $\rightarrow$ & ${\rm H} + {\rm H} + {\rm H}$ &
$k_9 \simeq 6.5\times10^{-7} T^{-1/2} e^{-52000/T}\times$\\
& & & $\qquad\qquad\times(1-e^{-6000/T}) \;{\rm cm^6\,s^{-1}}$\\

${\rm H} + {\rm H} + {\rm H}_2$ & $\rightarrow$ & ${\rm H}_2 + {\rm H}_2$ &
$k_{10} \simeq k_8/8$\\

${\rm H}_2 + {\rm H}_2$ & $\rightarrow$ & ${\rm H} + {\rm H} + {\rm H}_2$ &
$k_{11} \simeq k_9/8$\\

${\rm H} + e^-$ & $\rightarrow$ & ${\rm H}^+ + e^- + e^-$ &
$k_{\rm 12} \simeq 5.8\times10^{-11} T^{1/2}
	e^{-158000/T} \;{\rm cm^3\,s^{-1}}$\\

${\rm H} + {\rm H}$ & $\rightarrow$ & ${\rm H}^+ + e^- + {\rm H}$ &
$k_{\rm 13} \simeq 1.7\times10^{-4} k_{12}$\\

${\rm H}_2 + h\nu$ & $\rightarrow$ & ${\rm H}^+ + {\rm H}$ &
$k_{\rm 14} \simeq 1.1\times10^8 j(\nu_{\rm LW}) \; {\rm s^{-1}}$\\

${\rm D}^+ + e^-$ & $\rightarrow$ & ${\rm D} + h\nu$ &
$k_{\rm 15} \simeq 8.4\times10^{-11}
T^{-0.5} \times$\\
& & & $\quad\times({T\over{10^3}})^{-0.2}[1+({T\over{10^6}})^{0.7}]^{-1}
\; {\rm cm^3 s^{-1}}$\\

${\rm D} + {\rm H}^+$ & $\rightarrow$ & ${\rm D}^+ + {\rm H}$ &
$k_{\rm 16} \simeq 3.7\times10^{-10} T^{0.28} e^{-43/T}
\; {\rm cm^3 s^{-1}}$\\

${\rm D}^+ + {\rm H}$ & $\rightarrow$ & ${\rm D} + {\rm H}^+$ &
$k_{\rm 17} \simeq 3.7\times10^{-10} T^{0.28} \; {\rm cm^3 s^{-1}}$\\

${\rm D}^+ + {\rm H}_2$ & $\rightarrow$ & ${\rm H}^+ + {\rm HD}$ &
$k_{\rm 18} \simeq 2.1\times10^{-9} \; {\rm cm^3 s^{-1}}$\\

${\rm HD} + {\rm H}^+$ & $\rightarrow$ & ${\rm H}_2 + {\rm D}$ &
$k_{\rm 19} \simeq 1.0\times10^{-9} e^{-464/T} \; {\rm cm^3 s^{-1}}$\\
\hline\hline
\end{tabular}
\label{reaction_rates}
\end{table}

Since molecules play such an important role in the formation of the
first luminous objects, it is important to include a proper treatment of
their abundance evolution. However, a full treatment should keep into
account $\sim 20$ different species and $\sim 100$ different
reactions. For instance, GP98 give a chemical network
of 87 reactions, which includes $e^-$, H, H$^+$, H$^-$, D, D$^+$, He,
He$^+$, He$^{++}$, Li, Li$^+$, Li$^-$, \HH\index{HH}, \HHP, HD, HD$^+$, HeH$^+$,
LiH, LiH$^+$, H$_3^+$ and H$_2$D$^+$; even their {\it minimal model} is
too complicated to be described here, so we will just describe the most
basic processes involved in \HH\index{HH} formation. However, we note that the
papers by Abel \etal (1997)\footnote{The collisional rate coefficients
given in the Abel \etal (1997) paper can be readily obtained through a
FORTRAN code available on the web page
http://www.astro.psu.edu/users/tabel/PGas/LCA-CM.html ; also note that
on Tom Abel's web site (http://www.tomabel.com/) it is possible to find
several useful informations about the primordial universe in general.}\
and by GP98 provide a much more accurate description of
primordial chemistry\index{Chemistry}.

\subsubsection{Atomic Hydrogen\index{Hydrogen} and free electrons}
Apart from molecule formation (see below), the main reactions involving
Hydrogen (and Helium, to which we can apply all the arguments below) are
ionizations\index{Hydrogen!Ionization} and recombinations\index{Hydrogen!Recombination}.

Ionizations can be produced both by radiation (${\rm H} + h\nu
\rightarrow {\rm H}^+ + e^-$) and by collisions, mainly with free
electrons (${\rm H} + e^- \rightarrow {\rm H}^+ + 2e^-$) but also with
other H atoms (${\rm H} + {\rm H} \rightarrow {\rm H}^+ e^- + {\rm
H}$). Photoionizations dominate over collisions as long as UV photons
with energy above the H ionization\index{Hydrogen!Ionization} threshold are present (see \eg
Osterbrock 1989), but this is not always the case before the formation
of the first luminous objects, when only the CMB\index{Cosmic Microwave Background} radiation is present;
even after some sources of radiation have appeared, it is likely that
their influence will be mainly local, at least until reionization.
However, in the primordial universe the electron temperature is low,
and collisional ionizations are relatively rare.

Recombinations\index{Hydrogen!Recombination} (${\rm H}^+ + e^- \rightarrow {\rm H} + h\nu$) have much
higher specific rates, since they do not have a high energy
threshold. They probably dominate the evolution of free electrons, which
can be relatively abundant as residuals from the recombination era (\eg
Peebles 1993), and are important for \HH\index{HH} formation (see below).

Simple approximations for the reaction rates of collisional ionizations
and recombinations are given in Table \ref{reaction_rates}.


\subsubsection{\HH\index{HH} formation and disruption}
At present, \HH is commonly believed to form mainly through reactions
taking place on the surface of dust grains (but see \eg Cazaux \& Spaans
2004). Such a mechanism cannot work in the primordial universe, when the
metal (and dust) abundance was negligible. The first two mechanisms
leading to formation of \HH in a primordial environment was described by
McDowell (1961) and by Saslaw \& Zipoy (1967); soon after the
publication of this latter paper,
\HH\index{HH} started to be included in theories of structure formation (such as
the PD68 paper about globular cluster formation through \HH\index{HH cooling}
cooling). Both these mechanisms consist of two stages, involving either
$e^-$ or H$^+$ as catalyzers.  The first (and usually most important)
goes through the reactions
\begin{eqnarray}
{\rm H} + e^-       & \rightarrow & {\rm H}^- + h\nu\\
{\rm H}^- + {\rm H} & \rightarrow & {\rm H}_2- + e^-
\end{eqnarray}
while the second proceeds as
\begin{eqnarray}
{\rm H}^+ + {\rm H}   & \rightarrow & {\rm H}_2^+ + h\nu\\
{\rm H}_2^+ + {\rm H} & \rightarrow & {\rm H}_2- + {\rm H}^+.
\end{eqnarray}
In both cases, \HH\index{HH} production can fail at the intermediate stage if a
photodissociation occurs (${\rm H}^- + h\nu \rightarrow {\rm H} +
e^-$, or ${\rm H}_2^+ + h\nu \rightarrow {\rm H}^+ + {\rm H}$).
The rates of all these reactions are listed in Table \ref{reaction_rates}

By combining the reaction rates of these reactions, it is possible (see
Tegmark \etal 1997; hereafter T97) to obtain an approximate evolution for the
ionization fraction $x \equiv n_{\rm H^+} / n$ and the \HH fraction
$f_{\rm H_2} \equiv 2n_{\rm H_2}/n$ (here $n$ is the total density of
protons, $n \simeq n_{\rm H} + n_{\rm H^+} + 2n{\rm H_2}$):
\begin{eqnarray}
&& x(t)           \simeq {x_0\over{1+x_0 n k_1 t}}\\
&& f_{\rm H_2}(t) \simeq f_0 + 2{k_m\over k_1} \ln(1 + x_0 n k_1 t)
\qquad{\rm with}\\
&& k_m            =      {{k_2 k_3}\over{k_3+k_4/[n(1-x)]}} + 
                      {{k_5 k_6}\over{k_6+k_7/[n(1-x)]}}
\end{eqnarray}
where the various $k_i$ are the reaction rates given in Table
\ref{reaction_rates}, and $x_0$ and $f_0$ are the initial fractions of
ionized atoms and of \HH\index{HH} molecules.

Another mechanism for \HH formation, which becomes important at
(cosmologically) very high densities $n_{\rm H}\gsim10^8\;{\rm cm^-3}$
are the so-called 3-body reactions described by Palla \etal (1983).
While the previous mechanisms are limited by the small abundance of free
electrons (or of H$^+$), reactions such as
\begin{equation}
{\rm H} + {\rm H} + {\rm H} \rightarrow {\rm H}_2 + {\rm H}
\end{equation}
can rapidly convert all the hydrogen\index{Hydrogen} into molecular form, provided the
density is high enough. For this reason, they are likely play an
important role during the formation of a primordial protostar.

Finally, \HH\index{HH} can be dissociated through collisional processes such as
reactions 9 in table \ref{reaction_rates}, but probably the most
important process is its photo-dissociation by photons in the
Lyman-Werner bands (11.26-13.6 eV; but Abel \etal 1997 found that the
most important range is between 12.24 and 13.51 eV). These photons are
below the H ionization\index{Hydrogen!Ionization} threshold, therefore they can diffuse to large
distances from their source. So, any primordial source emitting a
relevant number of UV photons (\eg a $\sim100\msunabel$ star) is likely to
have a major feedback\index{feedback} effect, since it can strongly reduce the amount of
\HH\index{HH} in the halo where it formed (and also in neighbouring halos),
inhibiting further star formation.


\subsubsection{Approximate predictions}

\begin{figure}[t]
\epsfig{file=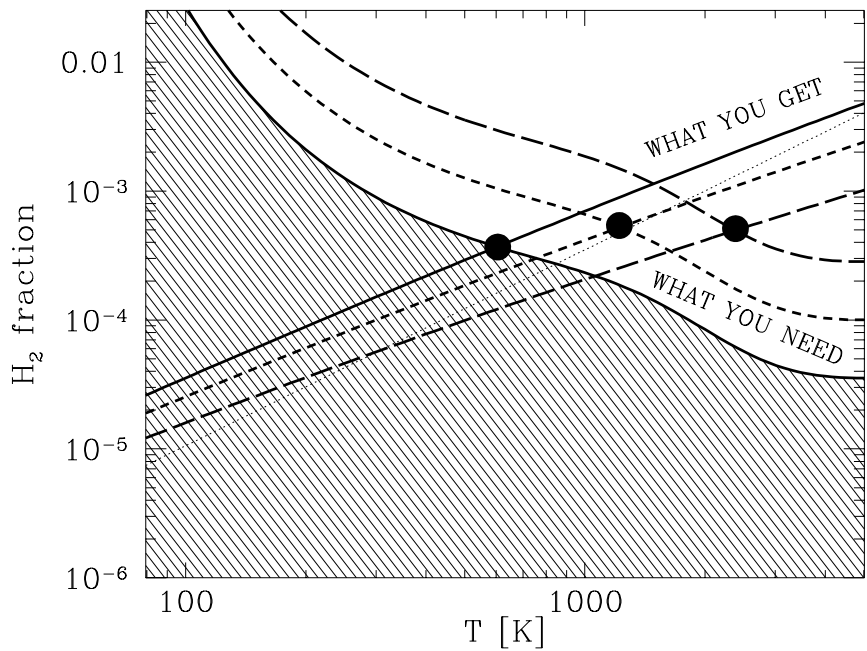,width=9truecm,angle=270}
\caption{Comparison of the \HH\index{HH} fraction needed for an halo to collapse
and \HH fraction which can be formed inside the halo in an Hubble time
(from T97). This is shown as a function of halo virial
temperature and for three different virialization redshifts ($z=100$:
solid; $z=50$: short dashes; $z=25$:long dashes). The three dots mark the
minimum \HH\index{HH} abundance which is needed for collapse at the three
considered redshift, and it can be seen that they all are at $f_{\rm
H_2}\sim 5\times10^{-4}$.}
\label{tegmark_minh2}
\end{figure}

\begin{figure}[t]
\epsfig{file=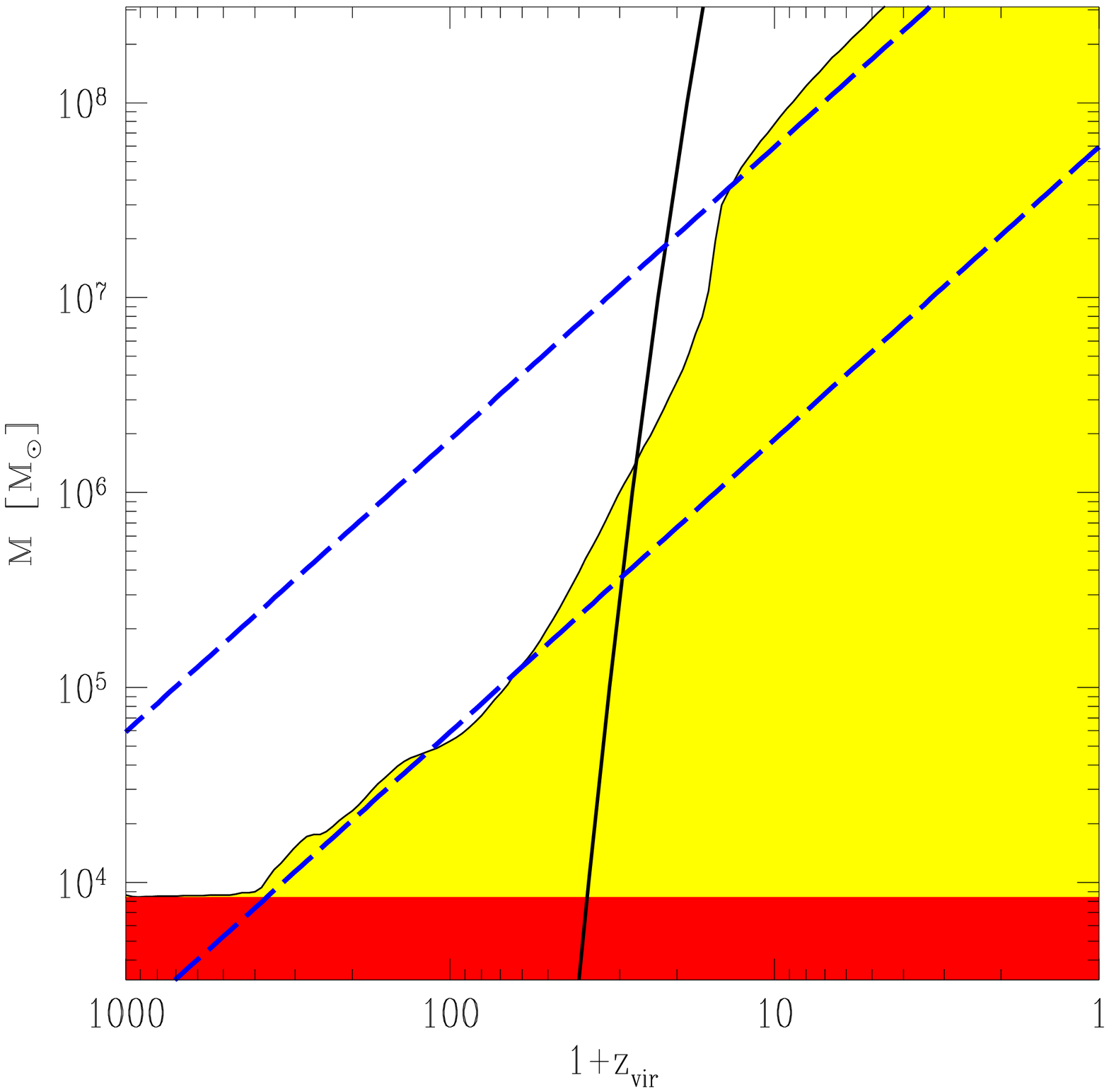,width=9truecm}
\caption{Evolution with virialization redshift of the minimum mass\index{minimum mass} which
is able to cool, collapse and possibly form stars. Only the halos whose ($z_{\rm
vir},M$) fall outside the filled region are able to cool fast
enough. The filled region is made of a red part, where CMB\index{Cosmic Microwave Background} radiation
prevents the cooling, and a yellow part where the cooling is slow
because of a dearth of \HH molecules. The two parallel dashed lines
correspond to virial temperatures of $10^4\;{\rm K}$ (the highest one)
and $10^3\;{\rm K}$ (the lowest one), while the almost vertical line in
the middle of the plot corresponds to $3\sigma$ peaks in a SCDM
($\Omega_{\rm m}=1,\ \Omega_\Lambda=0$) cosmology.}
\label{tegmark_minmass}
\end{figure}

The above information can be used for making approximate predictions
about the properties of the halos hosting the first luminous
objects. Such kind of predictions was started by Couchman \& Rees
(1986), and especially by T97 and their example was followed and
improved by several authors (see below).

The basic idea is to have a simplified model for the evolution of the
\HH\index{HH} fraction and the cooling rate inside spherical top-hat fluctuations,
in order to check whether after virialization the baryons are able to
cool (and keep collapsing), or they just ``settle down'' at about the
virial radius. Such approximate models are fast to compute, and it is
easy to explore the gas behaviour on a large range of virialization
redshifts and fluctuation masses (or, equivalently, virial temperatures).
In this way, for each virialization redshift it is possible to obtain
the minimum molecular fraction which is needed for the collapse to
proceed, and the minimum size of halos where such abundance is achieved.
The results interestingly point out that the molecular fraction
threshold separating collapsing and non-collapsing objects has an almost
redshift-independent value of $f_{\rm H_2}\sim5\times10^{-4}$ (see
fig. \ref{tegmark_minh2}). Instead, the minimum halo mass actually
evolves with redshift (see fig. \ref{tegmark_minmass}).

Predictions about the ability of the baryons inside each kind of halo to
keep collapsing after virialization can then be combined with
Press-Schechter\index{Press-Schechter Formalism} predictions about the actual abundances of halos.  For
instance, in fig. \ref{tegmark_minmass}\ the solid black (almost
vertical) line shows where the masses of $3\sigma$ fluctuations lie, as
a function of redshift. So, if we decide to neglect the rare
fluctuations at more than $3\sigma$ from the average, that figure tells
us that the first luminous objects can start forming only at $z\lsim30$,
in objects with a total mass $\gsim 2\times 10^6\;\msunabel$.

Such result is subject to a number of uncertainties, both about the
``details'' of the model and about the processes it neglects; here are
some of the more interesting developments:
\begin{enumerate}
\item{Abel \etal (1998) found that the minimum mass\index{minimum mass} is strongly
  affected by the uncertainties in the adopted \HH\index{HH cooling} cooling function,
  with differences that could reach a factor $\sim 10$ in the minimum
  mass\index{minimum mass} estimate}
\item{Fuller \& Couchman (2000) used numerical simulations\index{Numerical Simulation} of single,
  high-$\sigma$ density peaks in order to improve the spherical
  collapse approximation}
\item{Machacek \etal (2001) and Kitayama \etal (2001) investigated the
  influence of background radiation}
\item {Yoshida \etal (2003) used larger-scale numerical simulations\index{Numerical Simulation}
  and found that also the merging history of an halo could play a
  role, since frequent mergings heat the gas and prevent or delay the
  collapse of $\sim 30\%$ of the halos.}
\end{enumerate}

As a result of the improved modeling (and of a different set of
cosmological parameters as $\Lambda$CDM has substituted SCDM), in the
most recent papers the value of the minimum halo mass for the formation
of the first luminous objects is somewhat reduced to the range 0.5--1
$\times 10^6\; \msunabel$, with a weak dependence on redshift and a stronger
dependence on other parameters (background radiation, merging history
etc.).


\section{Numerical cosmological hydrodynamics}

Numerical simulations\index{Numerical Simulation} are an important tool for a large range of
astrophysical problems. This is especially true for the study of
primordial stars, given the absence of direct observational data about
these stars, and the relatively scarce indirect evidence we have.

\begin{figure}[p]
\epsfig{file=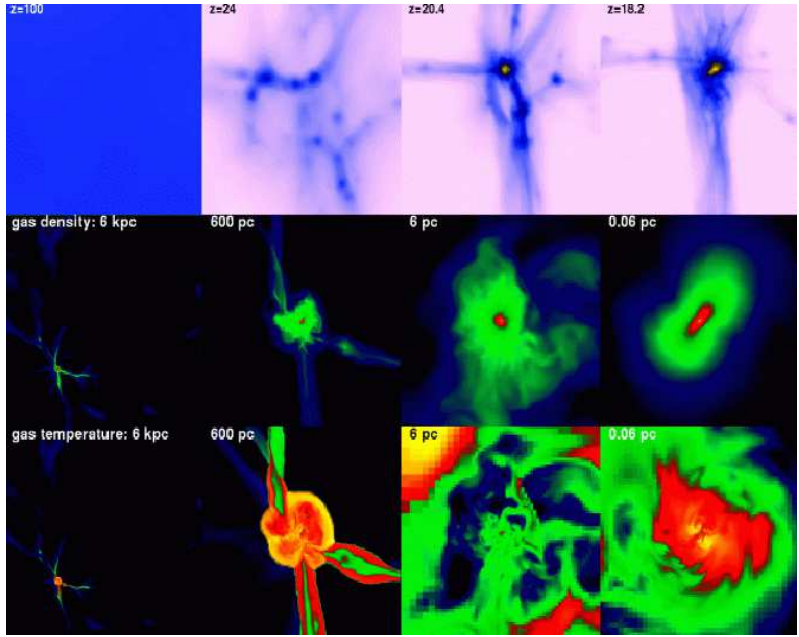,width=10.5truecm}
\caption{Overview of the evolution leading to the formation of a
primordial star. The top row shows the gas density, centered at the
pre-galactic object within which the star is formed. The four projections
are labelled with their redshifts. Pre-galactic objects form from very
small density fluctuations and continuously merge to form larger
objects. The middle and bottom rows show thin slices through the gas
density and temperature at the final simulation stage. The four pairs of
slices are labelled with the scale on which they were taken, starting
from 6 (proper) kpc (the size of the simulated volume) and zooming in
down to 0.06 pc (12,000 AU).  In the left panels, the larger scale
structures of filaments and sheets are seen. At their intersections, a
pre-galactic object of $\sim 10^6\;\msunabel$ is formed. The temperature
slice (second panel, bottom row) shows how the gas shock heats as it
falls into the pre-galactic object. After passing the accretion shock,
the material forms hydrogen\index{Hydrogen} molecules and starts to cool. The cooling
material accumulates at the centre of the object and forms the
high-redshift analog to a molecular cloud (third panel from the right),
which is dense and cold ($T\sim 200 K$). Deep within the molecular
cloud, a core of $\sim 100\;\msunabel$, a few hundred K warmer, is formed
(right panel) within which a $\sim 1\;\msunabel$ fully molecular object is
formed (yellow region in the right panel of the middle row). (from
ABN02).}
\label{halo_3d}
\end{figure}


\subsection{Adaptive refinement codes (ENZO)}
The two problems which immediately emerge when setting up a simulation
of primordial star formation are the dynamical range
and the required accuracy in solving the hydrodynamical equations.

When studying the formation of objects in a cosmological context, we
need both to simulate a large enough volume (with a box size of at least 100
comoving kpc), and to resolve objects of the size of a star
($\sim 10^{11}\;{\rm cm}$ in the case of the Sun), about 11 orders of
magnitude smaller.

This huge difference in the relevant scales of the problem is obviously
a problem. It can be attenuated by the use of Smoothed Particle
Hydrodynamics\index{Numerical Simulations!Smoothed Particle Hydrodynamics} (SPH; see \eg Monaghan 1992), whose Lagrangian nature has
some benign effects, as the simulated particles are likely to
concentrate (\ie, provide resolution) in the regions where the maximum
resolution is needed. However, even if this kind of method can actually
be employed (see Bromm \etal 1999, 2002), it has at least two important
drawbacks. First of all, the positive effects mentioned above cannot
bridge in a completely satisfactory way the extreme dynamical range we
just mentioned, since the mass resolution is normally fixed once and for
all at the beginning of the simulation. Second, SPH\index{Numerical Simulations!Smoothed Particle Hydrodynamics} is known to have
poor shock resolution properties, which casts doubts on the results when
the hydrodynamics becomes important.

The best presently available solution for satisfying both requirements
is the combination of an Eulerian approach (which is good for the
hydrodynamical part) and an adaptive technique (which can extend the
dynamical range). This is known as Adaptive Mesh Refinement\index{Numerical Simulations!Adaptive Mesh Refinement} (AMR; see
\eg Berger \& Colella 1989; Norman 2004), and basically consists in
having the simulation volume represented by a hierarchy of nested {\it
grids} ({\it meshes}) which are created according to resolution needs.

In particular, the simulations we are going to describe in the following
paragraphs were made with the code ENZO\footnote{The ENZO code can be
retrieved at the web site http://cosmos.ucsd.edu/enzo/} (see O'Shea
\etal 2004 and references therein for a full description). Briefly, ENZO
includes the treatment of gravitational physics through N-body
techniques, and the treatment of hydrodynamics through the piecewise
parabolic method (PPM) of Woodward \& Colella 1984 (as modified by Bryan
\etal 1995 in order to adapt to cosmological simulations).  ENZO can
optionally include the treatment of gas cooling (in particular \HH\index{HH
cooling} cooling, as described in the preceeding sections), of primordial
non-equilibrium chemistry\index{Chemistry} (see the preceeding sections) and of UV
background models (\eg the ones by Haardt \& Madau 1996) and heuristic
prescriptions (see Cen \& Ostriker 1992) for star formation in
larger-scale simulations.

\subsection{Formation of the first star}

\begin{figure}[p]
\epsfig{file=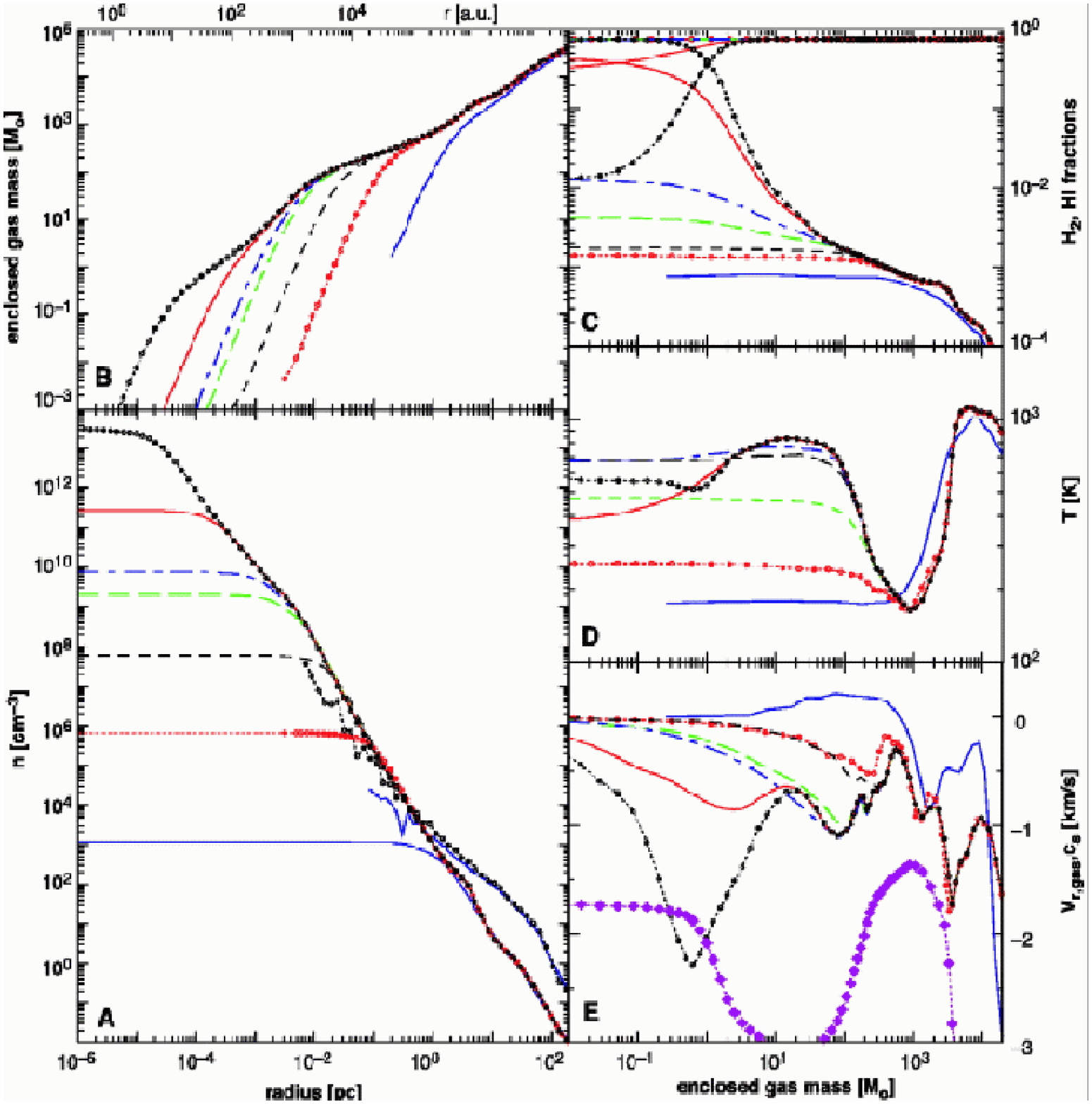,width=11.5truecm}
\caption{Radial mass-weighted averages of physical quantities at seven
different simulation times. (A) Particle number density in ${\rm
cm^{-3}}$ as a function of radius; the bottom line corresponds to
$z=19$, and moving upwards the ``steps'' from one line to the next are
of $9\times10^6$ yr, $3\times10^5$ yr, $3\times10^4$ yr, 3000 yr, 1500
yr, and 200 yr, respectively; the uppermost line shows the simulation
final state, at $z=18.181164$. The two lines between 0.01 and 200 pc
give the DM mass density (in ${\rm GeV\,cm^{-3}}$) at $z=19$ and the
final time, respectively.  (B) Enclosed gas mass.  (C) Mass fractions of
$H$ and \HH\index{HH}.  (D) Temperature. (E) Radial velocity of the baryons; the
bottom line in (E) shows the negative value of the local speed of sound
at the final time. In all panels the same output times correspond to the
same line styles. (from ABN02).}
\label{profiles3d}
\end{figure}

The use of AMR\index{Numerical Simulations!Adaptive Mesh Refinement} codes for cosmological simulations of the formation of
the first objects in the universe was pioneered by Abel \etal (2000) and
further refined by Abel \etal (2002; hereafter ABN02), where the dynamic
range covered by the simulations was larger than 5 orders of magnitude
in scale length, \ie the wide range between an almost cosmological
($\gsim 100$ comoving kpc, \ie $\gsim 5$ proper kpc) and an almost
stellar scale ($\lsim 1000\;{\rm AU} \sim 10^{-2}\;{\rm pc}$). In the
whole process, the AMR\index{Numerical Simulations!Adaptive Mesh Refinement} code kept introducing new (finer resolution)
meshes whenever the density exceeded some thresholds, or the Jeans
length\index{Jeans Length} was resolved by less than 64 grid cells.

The simulations were started at a redshift
$z=100$ from cosmologically consistent initial conditions\footnote{
Such conditions were taken from an SCDM model with $\Omega_\Lambda=0$,
$\Omega_{\rm m}=1$, $\Omega_{\rm b}=0.06$, $H_0=50\;{\rm
km/s\,Mpc^{-1}}$ which is quite different from the ``concordance''
$\Lambda$CDM model. However, the final results are believed to be only
marginally affected by differences in these cosmological
parameters.}, and the code also followed the non-equilibrium chemistry\index{Chemistry} of the
primordial gas, and included an optically thin treatment of radiative
losses from atomic and molecular lines, and from Compton cooling.

The main limitation of these simulations was the assumption that the
cooling proceeds in the optically thin limit; such assumption breaks
down when the optical depth inside \HH\index{HH} lines reaches unity
(corresponding to a Jeans length\index{Jeans Length} $\sim10^3\;{\rm AU}\sim 0.01\;{\rm
pc}$). However, the simulations were halted only when the optical
depth at line centres becomes larger than 10 (Jeans length\index{Jeans Length} of about
$\sim 10\;{\rm AU}$ ($\sim 10^{-4}\;{\rm pc}$), since it was unclear
whether Doppler shifts could delay the transition to the optically
thick regime.

\subsubsection{Summary of the evolution: radial profiles}
In Figures \ref{halo_3d} and \ref{profiles3d} we show the evolution of
gas properties both in pictures and in plots of spherically averaged
quantities, as presented in Abel \etal (2002).  From these figures,
and in particular from the local minima in the infall velocity
(Fig. \ref{profiles3d}e) it is possible to identify four
characteristic mass scales:
\begin{enumerate}
\item{The mass scale of the pre-galactic halo, of $\sim
7\times10^5\msunabel$ in total, consistent with the approximate
predictions discussed in the previous sections}
\item{The mass scale of a ``primordial molecular cloud'', $\sim
4000\msunabel$: the molecular fraction in this region is actually very low
($\lsim10^{-3}$), but it is enough to reduce the gas temperature from
the virial value ($\gsim10^3\;{\rm K}$) to $\sim 200\;{\rm K}$}
\item{The mass scale of a ``fragment'', $\sim 100\msunabel$, which is
determined by the change in the \HH cooling\index{HH cooling} properties at a density
$n\sim10^4\;{\rm cm^{-3}}$ (for a complete discussion, see Bromm \etal
1999, 2002), when Local Thermal Equilibrium is reached and the cooling
rate dependence on density flattens to $\Lambda\propto n$ (from
$\Lambda\propto n^2$); this is also the first mass scale where the gas
mass exceeds the Bonnor-Ebert mass (Ebert 1955, Bonnor 1956) $M_{\rm
BE}(T,n,\mu)\simeq 61\msunabel T^{3/2} n^{-1/2} \mu^{-2}$ (with $T$ in
Kelvin and $n$ in ${\rm cm^{-3}}$), indicating an unstable collapse}
\item{The mass scale of the ``molecular core'', $\sim 1\msunabel$ is
determined by the onset of 3-body reactions at densities in the range
$n\sim10^8-10^{10}\;{\rm cm^{-3}}$, which leads to the complete
conversion of hydrogen\index{Hydrogen} into molecular form; at this stage, the infall
flow becomes supersonic (which is the precondition for the appearance
of an accretion shock and of a central hydrostatic flow); the increase
in the \HH\index{HH} abundance due to the formation of this molecular core also
leads to the transition to the optically thick regime.}
\end{enumerate}

\subsubsection{Angular momentum}

\begin{figure}[t]
\epsfig{file=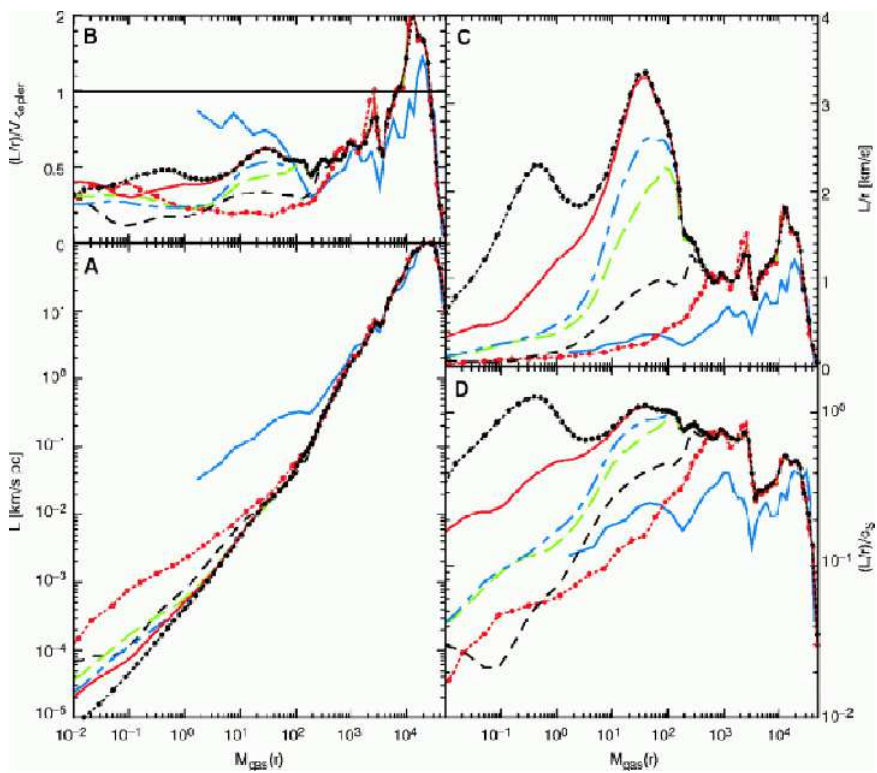,width=11.5truecm}
\caption{Radial mass weighted averages of angular momentum\index{Angular Momentum}-related
quantities at different times (the same as in
fig. \ref{profiles3d}). (A) specific angular momentum\index{Angular Momentum} $L$. (B)
Rotational speed in units of Keplerian velocity $v_{\rm
Kep}\equiv(GM_r/r)^{1/2}$. (C) Rotational speed ($L/r$). (D)
Rotational speed in units of the sound speed $c_s$ (from ABN02).}
\label{angmom3d}
\end{figure}

In Fig. \ref{angmom3d} we show the evolution of the radial distribution
of average specific angular momentum\index{Angular Momentum} (and related quantities).  It is
remarkable that in ABN02 simulations rotational support does not halt
the collapse at any stage, even if this could be a natural expectation.

There are two reasons for such (apparently odd) fact:
\begin{enumerate}
\item{As can be seen in panel A of Fig. \ref{angmom3d}, the collapse
starts with the central gas having much less specific angular momentum\index{Angular Momentum}
than the average (this is typical of halos produced by gravitational
collapse; see \eg Quinn \& Zurek 1988). That is, the gas in the central
regions starts with little angular momentum\index{Angular Momentum} to lose in the first place.}
\item{Second, some form of angular momentum\index{Angular Momentum} transport is clearly active,
as is demonstrated by the decrease in the central specific angular
momentum. Turbulence is likely to be the explanation: at any radius,
there will be both low and high angular momentum\index{Angular Momentum} material, and
redistribution will happen because of pressure forces or shock waves:
lower angular momentum\index{Angular Momentum} material will selectively sink inwards,
displacing higher angular momentum\index{Angular Momentum} gas. It is notable that this kind of
transport will be suppressed in situations in which the collapse occurs
on the dynamical time scale, rather than the longer cooling time scale:
this is the likely reason why this kind of mechanism has not been
observed in simulations of present day star formation (see \eg. Burkert \&
Bodenheimer 1993).}
\end{enumerate}

\subsection{SPH\index{Numerical Simulations!Smoothed Particle Hydrodynamics} results}

Bromm \etal (1999, 2002) have performed simulations of primordial
star formation using an SPH\index{Numerical Simulations!Smoothed Particle Hydrodynamics} code which included essentially the same
physical processes as the simulations we just described. Apart from the
numerical technique, the main differences were that they included
deuterium in some of their models, and that their initial conditions
were not fully cosmological (\eg, since they choose to study isolated
halos, the angular momenta are assigned as initial conditions, rather
than generated by tidal interactions with neighbouring halo).

\begin{figure}[t]
\epsfig{file=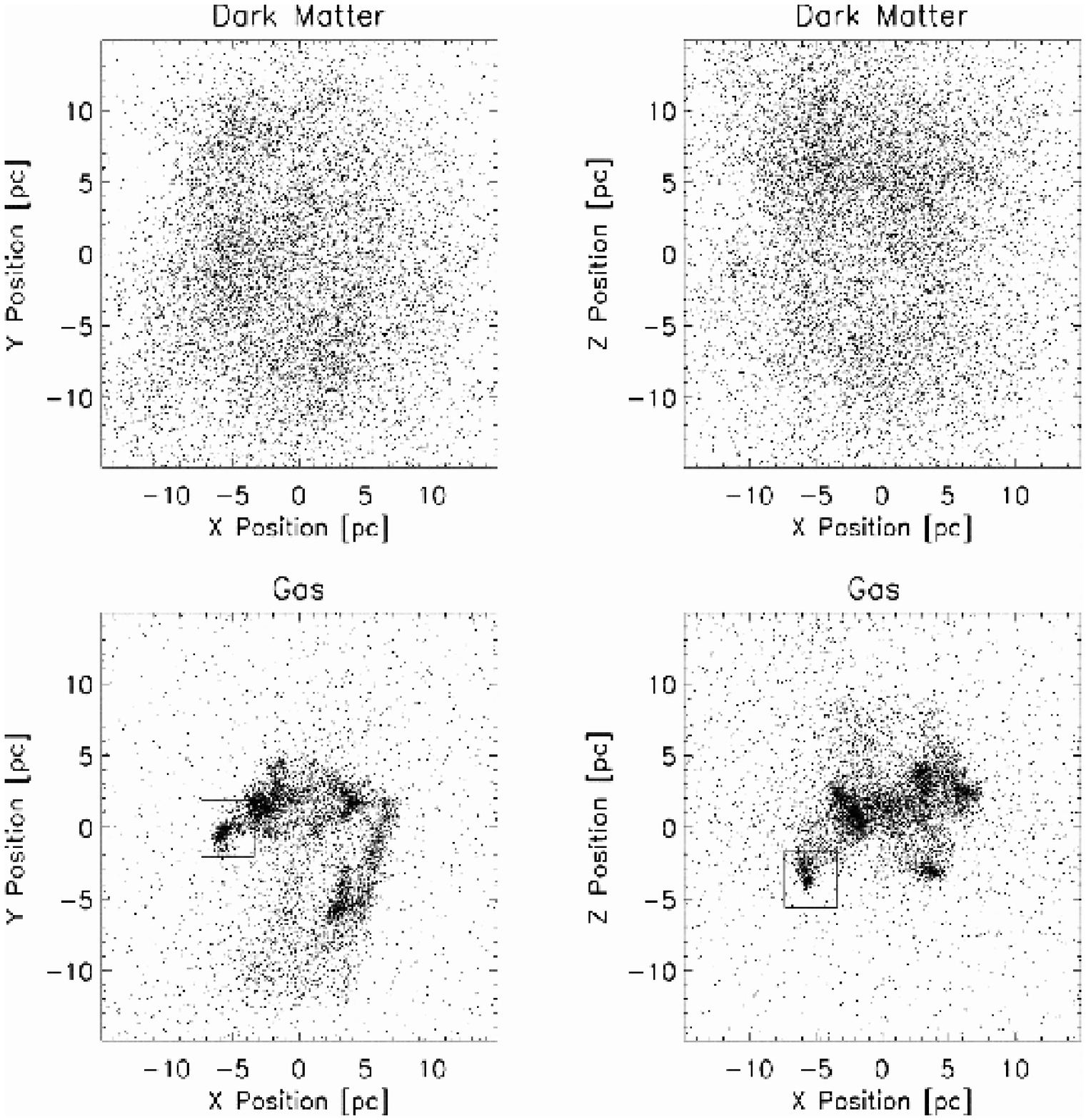,width=11.5truecm}
\caption{Typical result of the SPH\index{Numerical Simulations!Smoothed Particle Hydrodynamics} simulations by Bromm \etal 2002. The
figure shows the morphology of a simulated $2\times10^6\;\msunabel$ halo
virializing at $z\sim30$ just after the formation of the first clump of
mass $1400\;\msunabel$ (which is likely to produce stars). The two top row
panels shows the distribution of the dark matter\index{Dark Matter} (which is undergoing
violent relaxation). The two bottom panels show the distribution of gas,
which has developed a lumpy morphology and settled at the centre of the
potential well.}
\label{sphresults3d}
\end{figure}

\begin{figure}[t]
\epsfig{file=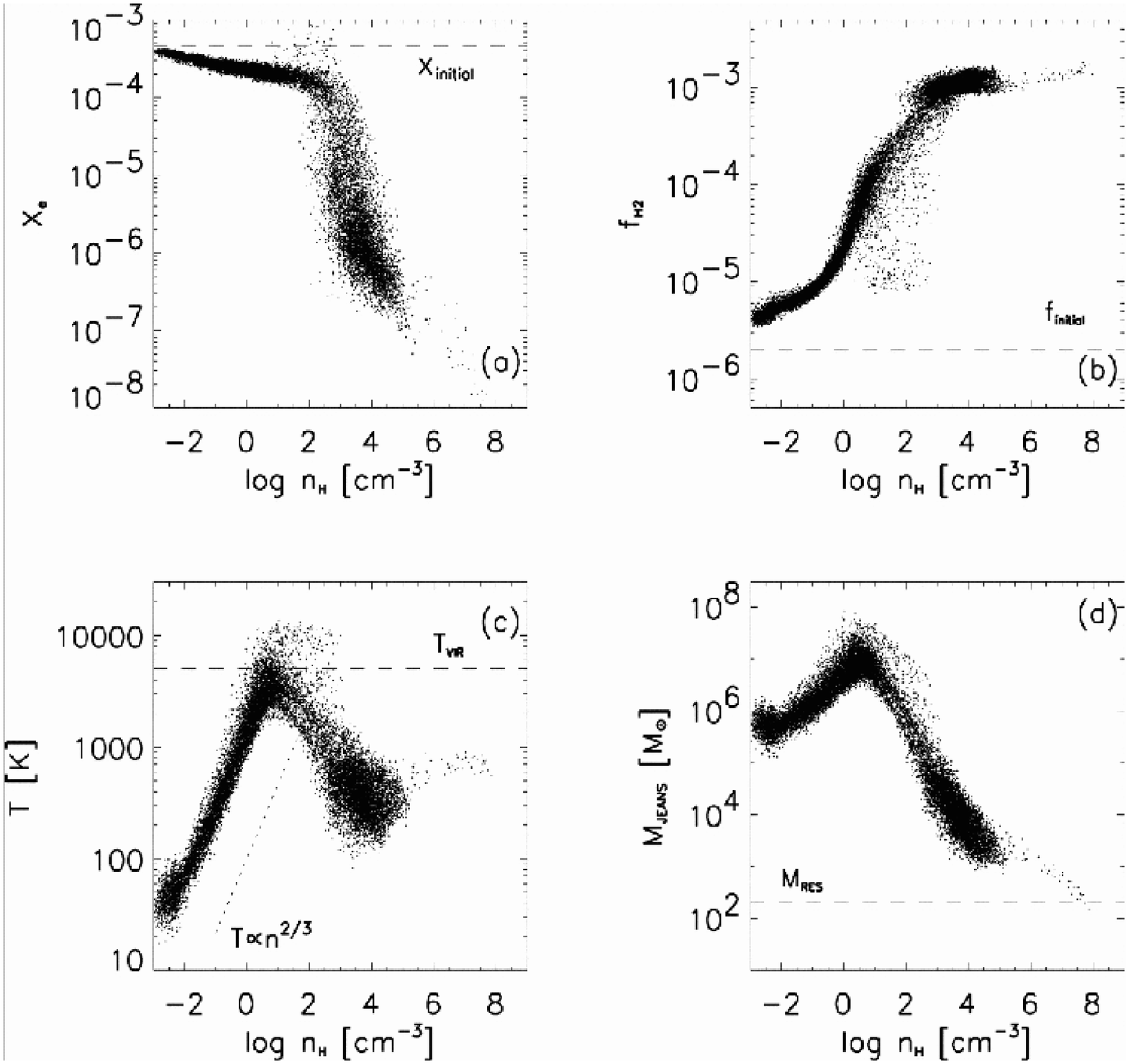,width=11.5truecm}
\caption{Gas properties in the SPH\index{Numerical Simulations!Smoothed Particle Hydrodynamics} simulations by Bromm \etal 2002. The
figure shows the properties of the particles shown in
Fig. \ref{sphresults3d}.  Panel (a) shows the free electron abundance
$x_e$ as a fraction of H number density. Panel (b) shows the \HH\index{HH}
abundance $f_{\rm H_2}$ (note that even the highest density particles
have $n_{\rm H}\lsim 10^8\;{\rm cm^3}$, so 3-body reactions are
unimportant and $f_{\rm H_2}\lsim10^{-3}$). Panel (c) shows the gas
temperature (low density gas gets to high temperatures because \HH
cooling\index{HH cooling}
is inefficient). Panel (d) shows the value of the Jeans mass\index{Jeans mass} which can
be obtained by using each particle temperature and density. All the
panels have the hydrogen\index{Hydrogen} number density $n_{rm H}$ on the X axis.}
\label{sphresultsgas}
\end{figure}

In Fig. \ref{sphresults3d} and \ref{sphresultsgas} we show the results
of one of their simulations, which give results in essential agreement
with those we have previously discussed.

An interesting extra-result of this kind of simulation is that the
authors are able to assess the mass evolution of the various gaseous
clumps in the hypothesis that feedback\index{feedback} is unimportant (which could be the
case if the clumps directly form intermediate mass \index{Black Hole}black holes without
emitting much radiation, or if fragmentation\index{fragmentation} to a quite small stellar
mass scale happens). They find that a significant fraction ($\sim 0.5$)
of the halo gas should end up inside one of the gas clumps, although it
is not clear at all whether this gas will form stars (or some other kind
of object). Furthermore, they find that clumps are likely to increase
their mass on a timescale of about $10^7\;{\rm yr}$ (roughly
corresponding to the initial time scale of the simulated halo), both
because of gas accretion and of mergers, and they could easily reach
masses $\gsim 10^4\;\msunabel$. Obviously, this result is heavily
dependent on the not very realistic assumed lack of feedback\index{feedback}.

\section{Protostar formation and accretion}

\begin{figure}[p]
\epsfig{file=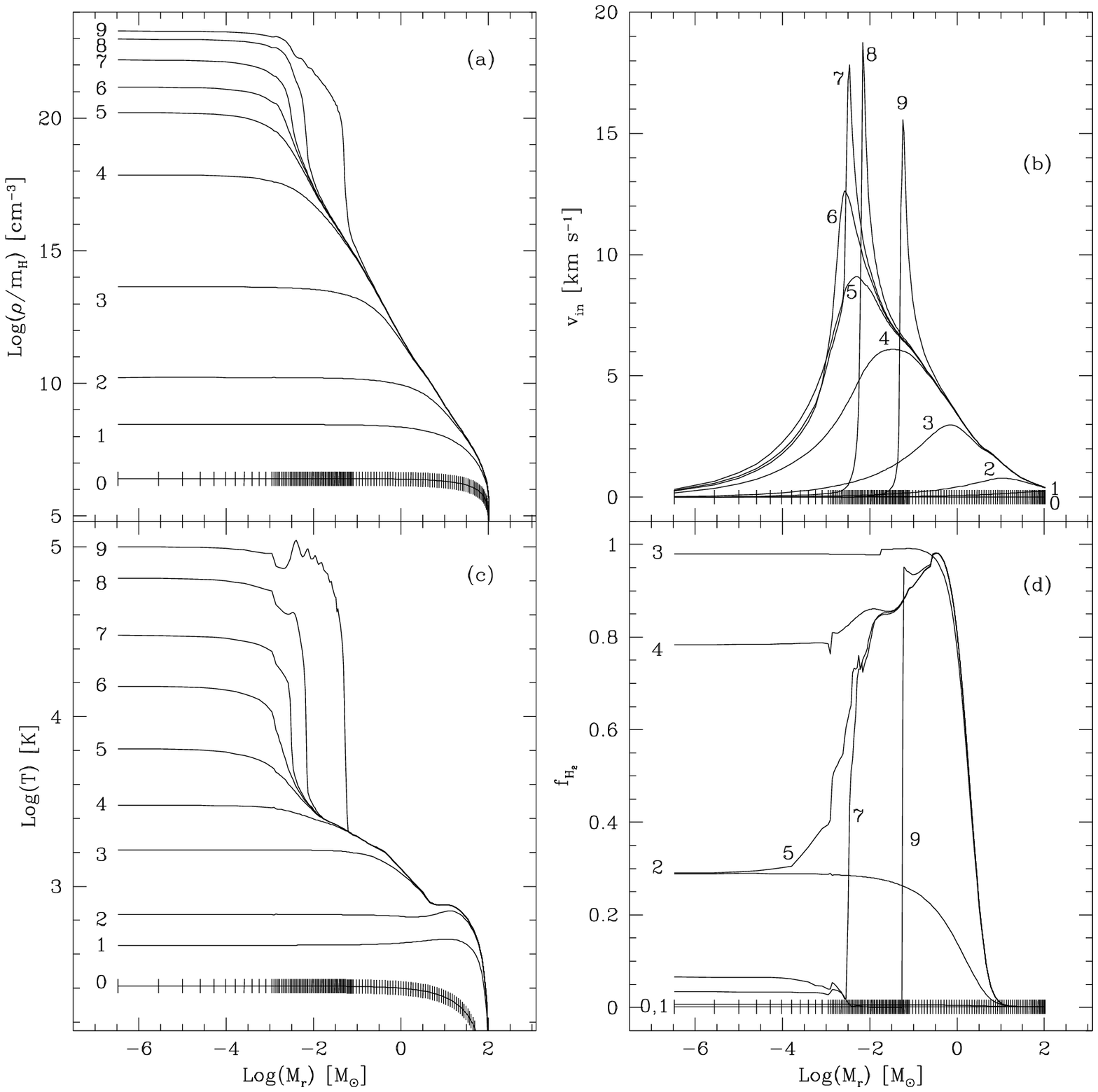,width=11truecm}
\caption{Proto-stellar collapse\index{protostellar collapse}, formation of an
hydrostatic core\index{protostellar collapse!hydrostatic core} and
start of the proto-stellar accretion\index{protostellar collapse!accretion} phase as can be found with 1-D
simulations. The four panel show the profiles of density (top left; a),
infall velocity (top right; b), temperature (bottom left;d) and \HH\index{HH}
abundance (bottom right; d) as a function of enclosed mass at 10
different evolutionary stages (0=initial conditions; 9=final stage of
the computation). The most relevant phases are the rapid formation of
\HH (2-3) and the formation of a shock on the surface of the hydrostatic
core (5-6-7), followed by the onset by accretion. Also note the almost
perfect power-law behaviour of the density profile before core
formation. (from Ripamonti \etal 2002).}
\label{protostellarcollapse}
\end{figure}

\begin{figure}[t]
\epsfig{file=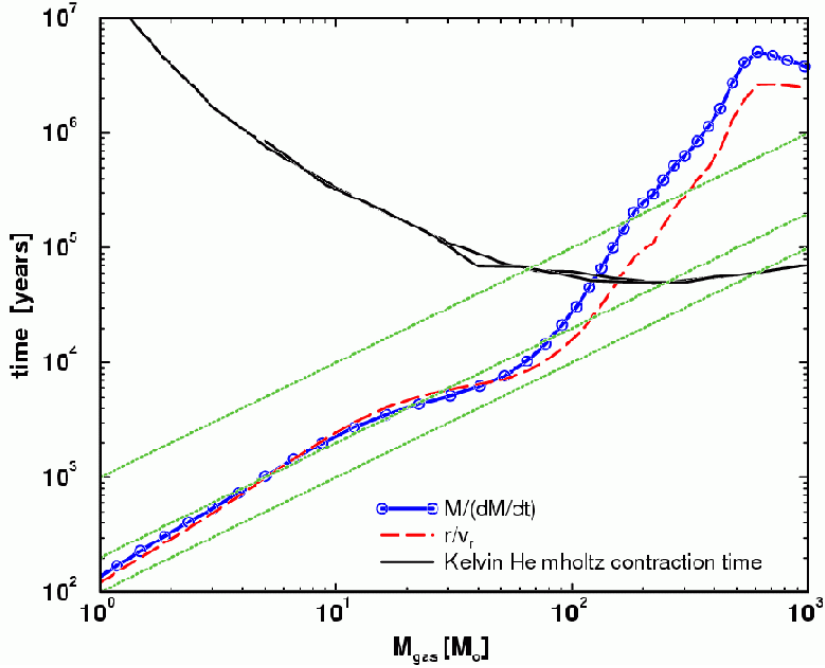,width=11truecm}
\caption{Comparison of the accretion\index{protostellar collapse!accretion} and the Kelvin-Helmoltz time scales
for a primordial protostars, as a function of protostellar mass. The
Kelvin-Helmoltz contraction time (obtained using the ZAMS luminosity as
given by Schaerer 2002) is shown as the solid black line, while
the dashed line and the solid line with circles show the time which is
needed to accrete each mass of gas; they are based on the results of
ABN02 and differ slightly in the way in which they were obtained. The
dotted lines mark three constant accretion rates of $10^{-2}$ (bottom line),
$5\times10^{-3}$ (middle line) and $10^{-3}\;{\rm \msunabel/yr}$ (top line).}
\label{kelvinhelmoltz}
\end{figure}

\begin{figure}[t]
\epsfig{file=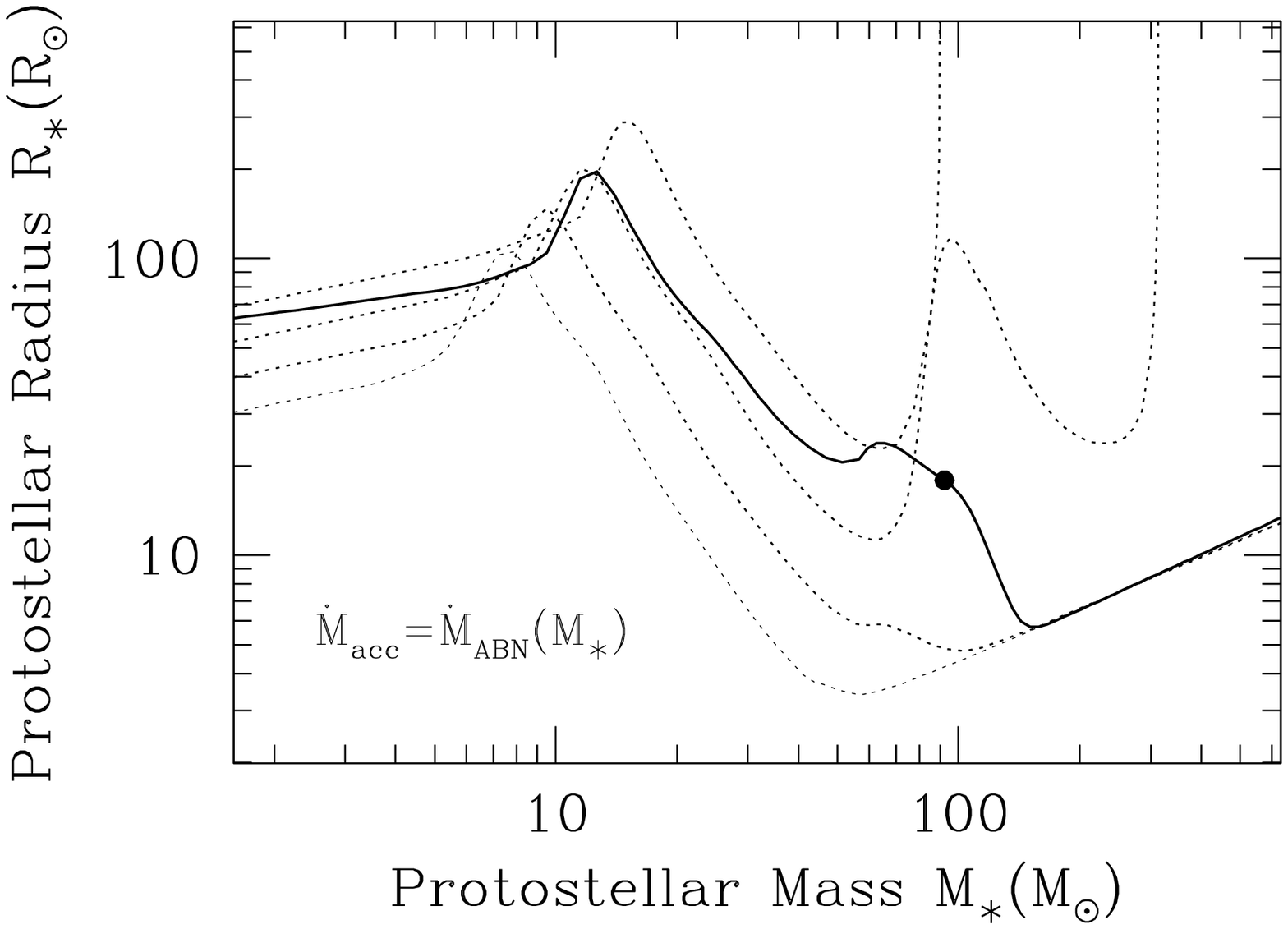,width=11truecm}
\caption{Evolution of the protostellar radius as a function of
protostellar mass in the models of Omukai \& Palla (2003). The models
differ in the assumed accretion rates. The dotted curves show the
results of models where $\dot{M}_{\rm core}$ is assumed to be constant
at values of $1.1\times 10^{-3}\;{\rm \msunabel/yr}$ (dotted curve starting
at $R_{\rm core}\simeq 30\;R_\odot$), $2.2\times 10^{-3}\;{\rm
\msunabel/yr}$ (starting at $R_{\rm core}\simeq 40\;R_\odot$), $4.4\times
10^{-3}\;{\rm \msunabel/yr}$ (``fiducial'' model; starting at $R_{\rm
core}\simeq 50\;R_\odot$) and $8.8\times 10^{-3}\;{\rm \msunabel/yr}$
(starting at $R_{\rm core}\simeq 65\;R_\odot$). The solid thick curve
shows the evolution of a model in which the accretion rate was obtained
from the extrapolation of the ABN02 data.}
\label{omukaipalla_accretion}
\end{figure}

Full three-dimensional simulations\index{Numerical Simulations!3D} (such as the ones of ABN02)
are not able to reach the stage when a star is really formed. They are
usually stopped at densities ($n\sim10^{10}-10^{\rm 11}\;{\rm cm^{-3}}$)
which are much lower than typical stellar densities ($\rho\sim1\; {\rm
g\,cm^{-3}}$, $n\sim10^{24}\; {\rm cm^{-3}}$).  In fact, at low
densities the gas is optically thin at all frequencies, and it is not
necessary to include radiative transfer in order to estimate the gas
cooling. Instead, at densities $n\gsim 10^{10}\; {\rm cm^{-3}}$ some of
the dominant \HH\index{HH} roto-vibrational lines become optically thick and
require the treatment of radiative transfer.

For this kind of problem, this is a prohibitive computational burden,
and at present the actual formation of a protostar can not be fully
investigated through self-consistent 3-D simulations\index{Numerical
Simulations!3D}. In order to
proceed further, it is necessary to introduce some kind of
simplification in the problem.

\subsection{Analytical results}
Historically, there were several studies based on analytical arguments
(\ie stability analysis\index{fragmentation!Stability Analysis}) or single-zone models, leading to different
conclusions about the properties of the final object.

An early example
is the paper by Yoneyama (1972), in which it was argued that
fragmentation\index{fragmentation!Opacity Limit} takes place until the opacity limit is reached. However,
Yoneyama looked at the opacity limit of the entire ``cloud'' (roughly
corresponding to one of the $10^5-10^6\;\msunabel$ mini-halos we consider at
present; originally, this was mass the scale proposed by PD68)
rather than the putative fragments and arrived at the conclusion that
fragmentation stopped for masses $\lsim 60\msunabel$.

More recently, Palla \etal 1983 pointed to the increase of the cooling rate at
$n\gsim 10^8\; {\rm cm^{-3}}$ (due to \HH\index{HH} fast formation through
3-body reactions) as a possible trigger for instability, leading to
fragmentation\index{fragmentation} on mass scales of $\sim 0.1\; \msunabel$; such instability
has actually been observed in the simulations of Abel \etal (2003),
but it does not lead to fragmentation because its growth is too slow
(see also Sabano \& Yoshii 1977, Silk 1983, Omukai \& Yoshii 2003 and
Ripamonti \& Abel 2004 for analytical fragmentation\index{fragmentation} criteria and their
application to this case).

\subsection{Mono-dimensional models}

\subsubsection{The formation of the protostellar core}
A less simplified approach relies on 1D studies\index{Numerical Simulations!1D}, such as those of Omukai
\& Nishi (1998) and Ripamonti \etal (2002): in such studies the heavy
price of assuming that the collapsing object always remains spherical
(which makes impossible to investigate the effects of angular momentum\index{Angular Momentum},
and prevents fragmentation\index{fragmentation}) is compensated by the ability to properly
treat all the other physical processes which are believed to be
important: first of all, the detailed radiative transfer of radiation
emitted in the numerous \HH\index{HH} resonance lines, then the transition to
continuum cooling (at first from molecular Collision-Induced Emission,
and later from atomic processes) and finally the non-ideal equation of
state which becomes necessary at high densities, when the gas becomes
increasingly hot and largely ionized.

Such studies find that the collapse initially proceeds in a
self-similar\index{protostellar collapse!self-similar phase} fashion (in good agreement with the solution found by Larson
1969 and Penston 1969); at the start of this phase (stages 1 and 2 in
fig. \ref{protostellarcollapse}, to be compared with the profiles at
corresponding densities in fig. \ref{profiles3d}) their results can be
compared with those of 3-D simulations\index{Numerical Simulations!3D}, and the agreement is good
despite the difference in the assumed initial conditions: this is likely
due to the self-similar properties. At later stages the comparison is
not as good, but this is presumably due to the differences in the
cooling rate (the 3D models at $n\gsim10^{10}\;{\rm cm^{-3}}$ only include
the optically thin cooling rate).

It is noticeable that the transition to optically thick cooling does not
stop the self-similar phase, contrary to theoretical expectations (see
\eg Rees 1976). The reason is that the reduction in the cooling rate is
smooth, and is further delayed by the onset of CIE\index{cooling!collision induced emission} cooling and by the
start of \HH\index{HH} dissociation, which acts as an heat well (each \HH
dissociation absorbs $4.48\;{\rm eV}$ of thermal energy) and provides an
alternative ``disposal'' mechanism for the excess thermal energy.

The self-similar phase\index{protostellar collapse!self-similar phase} proceeds until the central density is $n\gsim
10^{21}\;{\rm cm^{-3}}$ $\sim10^{-3}\;{\rm g\, cm^{-3}}$, when a small
($M_{\rm core,0}\sim 3\times 10^{-3}\; \msunabel$) hydrostatic
core\index{protostellar collapse!hydrostatic core}
eventually forms. Such core starts accreting the surrounding
gas at a very high rate (${\dot{M}}_{\rm core} \sim 0.01-0.1\; \msunabel\,
{\rm yr^{-1}}$), comparable to the theoretical predictions of Stahler
\etal 1980 
\begin{equation}
\dot{M}_{\rm core} \sim c_s^3/G \simeq
4 \times 10^{-3} \left({T\over{1000\;{\rm K}}}\right)^{3/2}\;
\left({\mu\over{1.27}}\right)^{-3/2}\; \msunabel\,{\rm yr^{-1}}
\end{equation}
where $c_s=[k_{\rm B}T/(\mu m_{\rm H})]^{1/2}$ is the isothermal sound
speed of a gas with mean molecular weight $\mu$.

Although the Omukai \& Nishi (1998) and the Ripamonti \etal
(2002) studies predict that such a high accretion rate will decline
(approximately as $\dot{M}_{\rm core}\propto t^{-0.3}$), it is clear
that, if the accretion proceeds unimpeded, it could lead to the
formation of a star with a mass comparable to the whole proto-stellar
cloud where the proto-star is born, that is, about $100\msunabel$; such a
process would take a quite short time of about $10^4-10^5\;{\rm
yr}$.

This scenario could be deeply affected by feedback\index{feedback} effects: even
not considering the radiation coming from the interior of the
proto-star, the energy released by the shocked material accreting onto
the surface can reach very high values, comparable with the Eddington
luminosity, and is likely to affect the accretion flow.

Figure \ref{kelvinhelmoltz}, compares the accretion
time scale (which can be extrapolated from the data at the end of the
ABN02 simulations) to the Kelvin-Helmoltz timescale,
\begin{equation}
t_{\rm KH} = {{GM^2}\over{RL}}
\end{equation}
where $L$ is the luminosity of the protostar. This plot tells us that
the accretion timescale is so fast that the stellar interior has very
little space for a ``re-adjustment'' (which could possibly stop the
accretion) before reaching a mass of $\sim10-100\msunabel$.

However, it can be argued that such a readjustment is not necessary,
since in the first stages most of the protostellar luminosity comes from
the accretion process itself: 
\begin{eqnarray}
&L_{\rm acc}\simeq {{GM_{\rm core}\dot{M}_{\rm core}}\over R_{\rm core}}
\nonumber \\ \simeq
8.5\times10^{37}
&\left({M_{\rm core}\over\msunabel}\right)
\left({\dot{M_{\rm core}}\over{0.01 \msunabel/yr}}\right)
\left({R_{\rm core}\over{10^{12}\;{\rm cm}}}\right)^{-1}
\;{\rm erg\,s^{-1}}
\end{eqnarray}
where we have inserted realistic values for $M_{\rm core}$, $\dot{M}_{\rm
core}$ and $R_{\rm core}$.
This luminosity is quite close to the Eddington luminosity
($L_{\rm Edd}\simeq 1.3\times10^{38}$ $(M/\msunabel) \;{\rm erg\,s^{-1}}$), and
radiation pressure could have a major effect.

Ripamonti \etal (2002) found that this luminosity is not
able to stop the accretion, but they do not properly trace the internal
structure of the core, so their results cannot be trusted except in the
very initial stages of accretion (say, when $M_{\rm core}\lsim0.1\msunabel$).

A better suited approach was used in studies by Stahler \etal (1986)
and, more recently, by Omukai \& Palla (2001, 2003), who assumed that
the accretion can be described as a sequence of steady-state accretion
flows onto a growing core. The core is modelled hydrostatically, as a
normal stellar interior (including the treatment of nuclear burning),
while the accreting envelope is modelled through the steady state
assumption, in conjunction with the condition that outside the
``photosphere'' (the region where the optical depth for a photon to
escape the cloud is $\gsim 1$) the accreting material is in
free-fall. As shown in Fig. \ref{omukaipalla_accretion}, Omukai \& Palla
(2003) find that even if feedback\index{feedback} luminosity deeply affects the
structure of the accreting envelope, it is never able to stop the
accretion before the proto-stellar mass reaches $60-100\;\msunabel$ as a
minimum; after that, the final mass of the protostar depends on the
assumed mass accretion rate $\dot{M}_{\rm core}$: for $\dot{M}_{\rm
core}\lsim 4\times 10^{-3}\;\msunabel\,{\rm yr^{-1}}$, the accretion can
proceed unimpeded until the gas is completely depleted (or the star
explodes as a supernova); otherwise, the radiation pressure is finally
able to revert the accretion; with high accretion rates this happens
sooner, and the final stellar mass is relatively low, while for
accretion rates only slightly above the critical value of $\sim 4\times
10^{-3}\;\msunabel\,{\rm yr^{-1}}$ the stellar mass can reach about
$300\;\msunabel$.

If such predictions are correct, the primordial Initial Mass Function\index{Initial Mass Function}
is likely to be much different from the present one, reflecting mainly
the mass spectrum of the gas fragments from which the stars originate,
and a mass of $\sim 100\;\msunabel$ could be typical for a primordial
star. However, we note that this important results could change as a
result of better modeling. For example, deep modifications of the envelope
structure are quite possible if a frequency-dependent opacity (rather than the
mean opacity used in the cited studies) is included.


\section{Discussion}
The previous sections show that, although not certain at all (because of
the big uncertainty about feedback\index{feedback} effects), numerical simulations\index{numerical simulation} tend
to favour the hypothesis that primordial stars had a larger typical mass
than present-day stars.

If this is true, it could indeed solve some observational puzzle, such
as why we have never observed a single zero-metallicity star (answer:
they have exploded as supernovae and/or transformed into compact objects
at a very high redshift), and maybe help explaining the relatively high
metallicities ($Z\gsim10^{-4}Z_\odot$ even in low column density
systems) measured in the Lyman $\alpha$ forest (see \eg Cowie \&
Songaila 1998), or the high ($\sim$ solar) metallicities we observe in
the spectra of some quasars already at redshift 6 (Fan \etal 2000, 2001,
2003, 2004; Walter \etal 2003).

However, a top-heavy primordial IMF also runs into a series of
problems, which we will discuss in the remaining of this section.

\subsection{UV Radiation feedback\index{feedback}}

\begin{figure}[t]
\epsfig{file=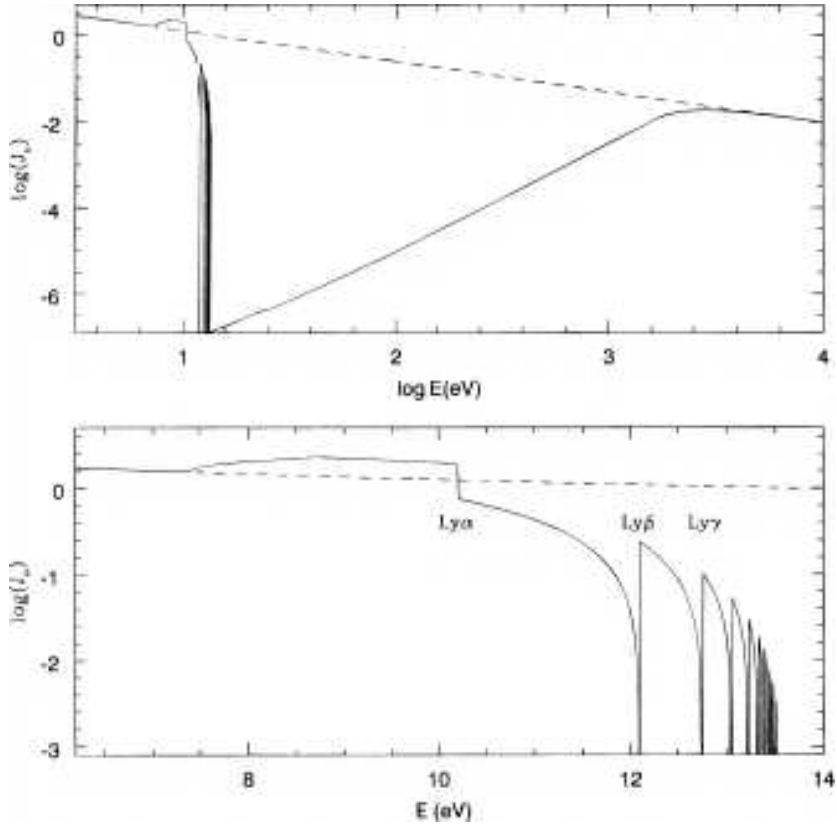,width=11truecm}
\caption{Average flux (in units of ${\rm erg\, cm^{-2}\, s^{-1}\,
sr^{-1}\, Hz^{-1}}$) at an observation redshift $z_{\rm obs}=25$ coming
from sources turning on at $z=35$.
The top panel shows the effects of absorption of neutral hydrogen\index{Hydrogen} and
helium, strongly reducing the flux between $\sim 13.6\;{\rm eV}$ and
$\sim 1\;{\rm keV}$ (the solid and dashed lines show the absorbed and
unabsorbed flux,  respectively).
The bottom panel shows the same quantities in a much smaller energy
range, in order to illustrate the sawtooth modulation due to line
absorption below 13.6 eV. (from Haiman, Rees \& Loeb 1997).}
\label{sawtooth}
\end{figure}

\begin{figure}[t]
\epsfig{file=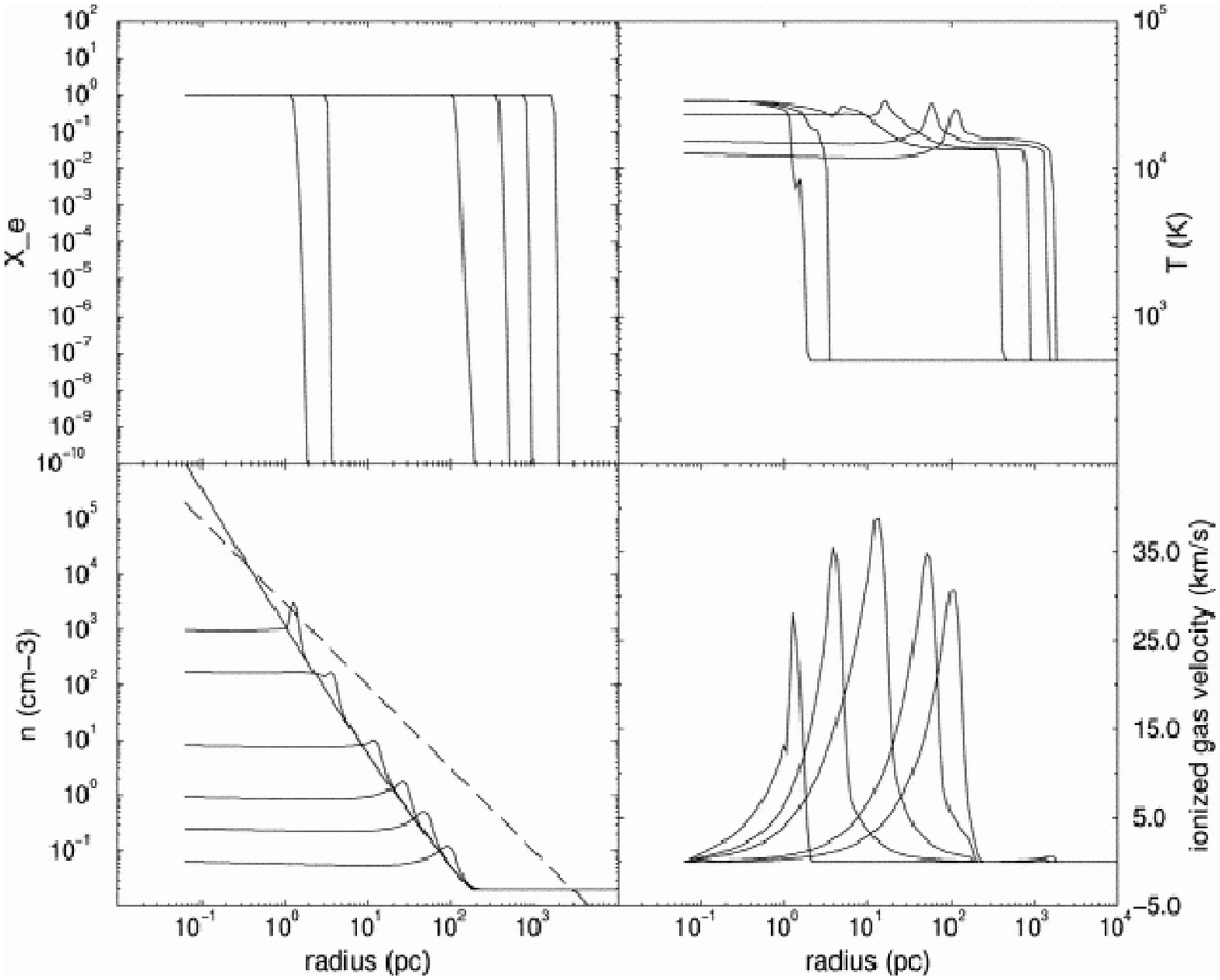,width=11truecm}
\caption{Dynamical evolution of the HII region produced by a
$200\,\msunabel$ star. Panels refer to ionization fraction profiles (top
left), temperature distributions (top right), densities (bottom left)
and velocities (bottom right).  The dashed line in the density panel is
for the Str\"omgren density (the density required to form a Str\"omgren
sphere and therefore initially bind the ionization front) at a given
radius. Profiles are given at times (from top to bottom in the density
panel; from left to right in the others) 0 (density panel), 63 kyr (all
panels), 82 kyr (ionization and temperature panels), 95 kyr (ionization
panel) 126.9 kyr (all panels), 317 kyr (all panels), 1.1 Myr
(temperature, density and velocity panels) and 2.2 Myr (all
panels), which is the approximate main sequence lifetime of a
$200\;\msunabel$ star. (From Whalen, Abel \& Norman 2004).}
\label{primordialhii}
\end{figure}

First of all, if a moderately massive star (even $M\gsim 20-30\msunabel$ is
likely to be enough) actually forms in the centre of an halo in the way
shown by ABN02, it will emit copious amounts of UV radiation, which will
produce important feedback\index{feedback} effects. Indeed, the scarcity of metals in
stellar atmospheres (see Schaerer 2002, and also Tumlinson \& Shull
2000, Bromm Kudritzki \& Loeb 2001) results in UV fluxes which are
significantly larger than for stars of the same mass but with
``normal'' metallicity. Furthermore, this same scarcity of metals is
likely to result in a negligibly small density of dust particles,
further advancing the UV fluxes at large distances from the sources when
compared to the present-day situation.

Massive primordial objects are also likely to end up into \index{Black Hole}black holes
(either directly or after having formed a star\footnote{Here we think of
a star as an object where quasi-hydrostatic equilibrium is reached
because of stable nuclear burning; according to this definition, a black
hole can be formed without passing through a truly stellar phase,
even if it is very likely to emit significant amounts of radiation in
the process.}), which could be the seeds of present day super-massive
\index{Black Hole}black holes (see \eg Volonteri, Haardt \& Madau 2003). If accretion can
take place onto these \index{Black Hole}black holes (also known as {\it mini-quasars}),
they are likely to emit an important amount of radiation in the UV and
the X-rays (this could also be important in explaining the WMAP result
about Thomson scattering optical depth; see \eg Madau \etal 2003).

UV photons can have a series of different effects. First of all,
we already mentioned in the section about chemistry\index{chemistry} that Lyman-Werner
photons ($11.26\;{\rm eV}\leq h\nu \leq 13.6\;{\rm eV}$) can dissociate
\HH\index{HH} molecules, preventing further star formation. Such photons are
effective even at large distances from their sources, and even in a
neutral universe, since their energy is below the H ionization\index{Hydrogen!ionization}
threshold; the only obstacle they find on their way are some of the
Lyman transitions of H, which can scatter these photons or remove them
from the band; this results in a ``sawtooth'' modulation (see
fig. \ref{sawtooth}) of the spectrum.
Haiman, Rees \& Loeb (1997) argue that this negative feedback\index{feedback} could
conceivably bring primordial star formation to an early stop; however,
other authors found that this is not
necessarily the case (Ciardi \etal 2000; Omukai 2001; Glover \& Brand
2001), or that the feedbac\index{feedback}k effect on
\HH\index{HH} abundance can be moderated by the effects of X-ray photons coming
from mini-quasars (Haiman Abel and Rees 2000; Glover \& Brand 2003).
 
A second obvious effect from UV photons is to ionize the interstellar
and the intergalactic medium; it is unclear whether these kind of
objects are important in the reionization history of the universe, but
they will definitely form HII regions in their immediate
neighbourhoods. In fig. \ref{primordialhii} we show the results of
Whalen, Abel \& Norman 2004 about the evolution of an HII region
produced by a $200\msunabel$ star inside an halo with the same density
profile as found in the ABN02 paper: in the beginning the ionization
front is D-type (that is, it is ``trapped'' because of the density) and
expands because a shock forms in front of it; later (after about 70 kyr
from the start of the UV emission) the ionization front reaches regions
of lower densities and becomes R-type (radiation driven), expanding much
faster than the shock. However, the shock keeps moving at speeds of the
order of $30\;{\rm km\,s^{-1}}$ even after the source of ionizing
photons is turned off. Such a speed is much larger than the rotational
velocities of the mini-halos where primordial star formation is supposed
to be taking place ($\lsim 10\;{\rm km s^{-1}}$), so this shocks are
likely to expel a very large fraction of the original gas content of the
mini-halo. UV emission could
even lead to the {\it photo-evaporation} of neighbouring
mini-halos, similar to what Barkana \& Loeb 1999 (see also Shapiro,
Iliev \& Raga 2004) found in the slightly different context to
cosmological reionization.

Since the dynamical timescale of a mini-halo is $\gsim 10^7\;{\rm yr}$
and is longer than the timescale for this kind of phenomena (definitely
shorter than the $\sim1-10$ Myr main sequence lifespan of massive stars,
and very likely to be of the order of $\sim 10^4-10^5\;{\rm yr}$), the
star formation in one mini-halo is likely to stop almost immediately
after the first (massive) object forms. This means that, unless two or
more stars form exactly at the same time (within $\sim1\%$ of the
mini-halo dynamical time), each mini-halo will form {\it exactly one}
massive star.

\subsection{Supernovae\index{supernovae} feedback and metallicities}

\begin{figure}[p]
\epsfig{file=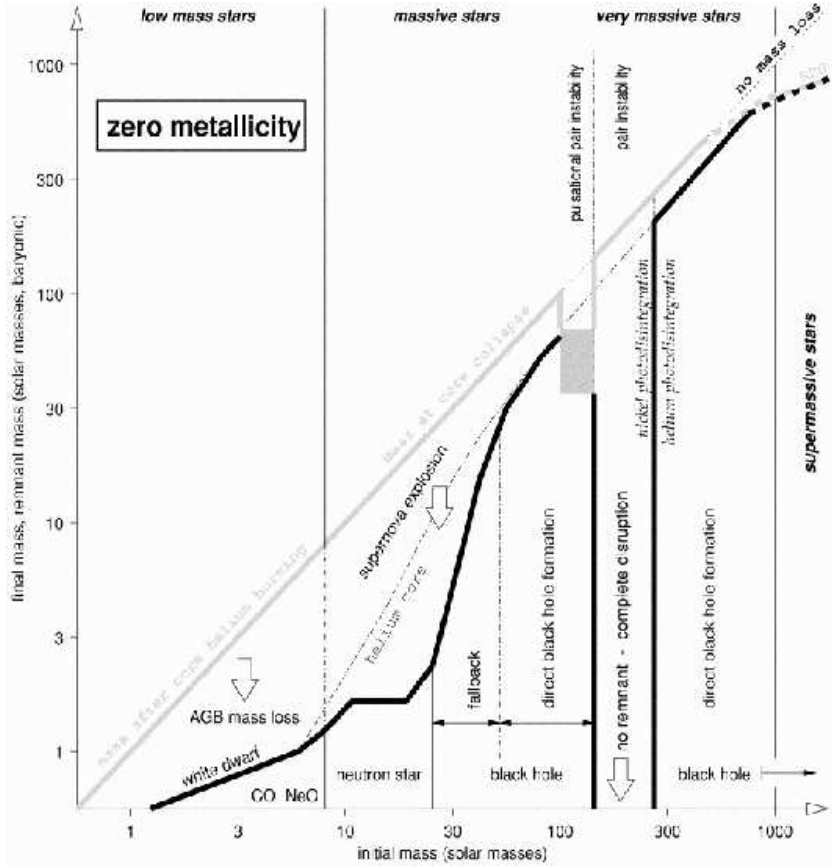,width=11truecm}
\caption{The ``final mass function'' of non-rotating metal-free stars as
a function of their initial mass, as given by from Heger \& Woosley
2002. The thick black curve gives the final mass of the collapsed
remnant, and the thick gray curve and the mass of the star at the start
of the event that produces that remnant (mass loss, SN etc.); for
zero-metal stars this happens to be mostly coincident with the dotted
line, corresponding to no mass loss. Below initial masses of $5-10\;\msunabel$
white dwarfs are formed; above that, initial masses of up to
$\sim25\;\msunabel$ lead to neutron stars. At larger masses, \index{Black Hole}black holes can
form in different ways (through fall-back of SN ejecta or
directly). Pair instability starts to appear above $\sim100\;\msunabel$, and
becomes very important at $\sim 140\;\msunabel$: stars with initial masses
in the $140-260\;\msunabel$ range are believed to completely disrupt the
star, leaving no remnant; instead, above this range pair instability is
believed to lead to a complete collapse of the star into a \index{Black Hole}black hole.}
\label{starfate}
\end{figure}

After producing plenty of UV photons during their life, primordial
massive stars are likely to explode as supernovae\index{supernovae}. This must be true for
some fraction of them, otherwise it would be impossible to pollute the
IGM with metals, as mass loss from zero-metal stars is believed to be
unimportant (this applies to stars with $M\lsim500\;\msunabel$; see Baraffe,
Heger \& Woosley 2001, and also Heger \etal 2002); however, it is clearly
possible that some fraction of primordial stars directly evolve into black
holes.

In figure \ref{starfate} we show the results about the fate of
zero metallicity stars, as obtained by Heger \& Woosley 2002 (but see
also Heger \etal 2003), indirectly confirming this picture and
suggesting that pair-instability supernovae\index{supernovae} could play a major role if
the primordial IMF is really skewed towards masses $\gsim
100\;\msunabel$.

\begin{figure}[t]
\epsfig{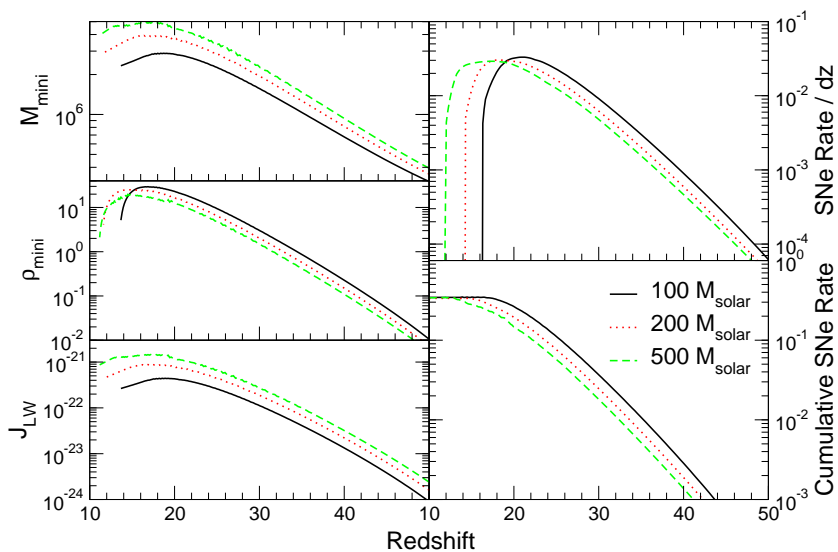}
\caption{Primordial supernova\index{supernovae} properties as reported by Wise \& Abel
2004. The right panels show the differential rate of primordial
supernovae per unit redshift (top) and the cumulative rate (bottom);
both are per year of observation and per square degree. The left panels
show the critical halo mass (in $\msunabel$) for primordial star formation
(top), the comoving number density (in ${\rm Mpc^{-3}}$) of halos above
the critical mass (middle) and the predicted specific intensity of the
soft UV background in the Lyman-Werner band (in ${\rm erg\; s^{-1}\;
cm^{-2}\; Hz^{-1}\; sr^{-1}}$, bottom). The three lines in each panel
refer to fixed primordial stellar masses of $100\msunabel$ (solid),
$200\msunabel$ (dotted) and $500\msunabel$ (dashed).}
\label{primordialsn}
\end{figure}

This has a series of consequences. First of all, supernova\index{supernovae} explosion are
one of the very few phenomena directly involving primordial stars which
we can realistically hope to observe. Wise \& Abel (2004; see also Marri
\& Ferrara 1998 and Marri, Ferrara \& Pozzetti 2000 for the effects of
gravitational lensing) investigated the expected number of such
supernovae by means of a semi-analytic model combining Press-Schechter
formalism\index{Press-Schechter formalism}, an evolving minimum mass\index{minimum mass} for star forming halos and negative
feedbacks, finding the rates shown in Fig. \ref{primordialsn}; if such
objects are pair-instability supernovae with masses $\gsim175\;\msunabel$,
they should be detectable by future missions such as JWST\footnote{James
Webb Space Telescope; more information at http://www.jwst.nasa.gov}; if
some of them are associated to Gamma Ray Bursts, the recently launched
{\it Swift}\footnote{More information at http://swift.gsfc.nasa.gov}
satellite has a slim chance of observing them at a redshifts
$\lsim30$ (see Gou \etal 2004).

Supernova\index{supernovae} explosions obviously have hydrodynamical effects on the gas of
the mini-halo where they presumably formed, especially in the case of
the particularly violent pair-instability supernovae (see \eg Ober, El
Elid \& Fricke 1983), and they are likely to expel it even if it was not
removed by the effects of UV radiation (see \eg MacLow \& Ferrara 1999
and Ferrara \& Tolstoy 2000), further reducing the probability of having
more than 1 star per mini-halo. The similarity with UV effects extends
to the influence on neighbouring halos which can be wiped out by SN
explosions (Sigward, Ferrara \& Scannapieco 2004).

Finally, supernovae provide a mechanism for spreading metals into the
IGM, as discussed \eg by Madau, Ferrara \& Rees 2001. In turn, these
metals will modify the cooling properties of the gas, and lead to star
formation with a normal (Salpeter-like) IMF.
If this is true, and if pair-instability supernovae\index{supernovae} dominate
the yields from primordial stars, we should be able to distinguish the
peculiar pair-supernova abundance pattern (described in Heger \& Woosley
2001) as a ``signature'' of primordial origin.

Searches for low-metallicity stars have a long history (see \eg Beers
1999, 2000), and their inability to find stars with metallicities
$Z\leq10^{-4}Z_\odot$ led to speculation that this metallicity marked
the transition to a normal IMF (see \eg Bromm \etal 2001, Schneider
\etal 2002).  However, Christlieb \etal 2002 finally reported the
discovery of an extremely metal poor star (HE0107-5240) with
$[Fe/H]=-5.3$\footnote{$[Fe/H] = \log_{10}(N_{Fe}/N_H) -
\log_{10}(N_{Fe}/N_H)_\odot$.}, a level compatible with a ``second
generation'' star (\ie, a star formed from material enriched by very few
supernovae; for comparison, Wise \& Abel 2004 find that primordial SNe
could enrich the IGM to $[Fe/H]\sim-4.1$, although this is probably an upper
limit). The abundance patterns in this star are quite strange (for
example, the carbon abundance is slightly less than 1/10 solar), but a
much better fit is provided by supernova\index{supernovae} yields of moderately massive
stars ($15-40\;\msunabel$) rather than from yields predicted for
pair-instability supernovae coming from very massive stars. However, at
the moment no model can satisfactorily fit all the observed abundances
(see Bessel, Christlieb \& Gustafsson 2004).

Even if the primordial nature of this star must still be established, it
represents a cautionary tale about the currently favoured predictions of
a large number of massive or very massive primordial stars, and a
reminder that better theoretical modeling is still needed, with
particular regard to feedback effects during the stellar accretion phase.

\section*{Acknowledgements}
We are grateful to the SIGRAV School for providing the opportunity to
write this lecture notes. We thank M. Mapelli and M. Colpi
for assistance and useful comments about the manuscript. This work was
started when both authors were at the Astronomy Department of
Pennsylvania State University. E.R. gratefully acknowledges support from
NSF CAREER award AST-0239709 from the U.S. National Science Foundation,
and from the Netherlands Organization for Scientific Research (NWO)
under project number 436016.


\References

\begin{thereferences}
\item Abel T, Anninos P, Zhang Y and Norman M 1997 {\it New Astr.} {\bf
2} 181
\item Abel T, Anninos P, Norman M L and Zhang Y 1998 {\it ApJ} {\bf 508} 518--529
\item Abel T, Bryan G L and Norman M L 2000 {\it ApJ} {\bf 540} 39
\item Abel T, Bryan G L and Norman M L 2002 {\it Science} {\bf 295}
  93--98 [ABN02]
\item Baraffe I, Heger A and Woosley S E 2001 {\it ApJ} {\bf 550} 890
\item Barkana R and Loeb A 1999 {\it ApJ} {\bf 523} 54
\item Barkana R and Loeb A 2001 {\it Physics Reports} {\bf 349}
125-238
\item Beers T C 1999 in {\it The Third Stromlo Symposium: The Galactic
  Halo}, eds. Gibson B K Axelrod T S \& Putman M E, {\it Astron. Soc.
  Pacif. Conf. Ser.} {\bf 165} 202
\item Beers T C 1999 {\it The First Stars}, proceedings
of the MPA/ESO Workshop held at Garching, Germany, 4-6 August 1999,
eds. A. Weiss, T. Abel \& V. Hill (Springer), p. 3
\item Bennett C L \etal 2003 {\it ApJS} {\bf 148} 1
\item Berger M J and Colella P 1989 {\it J. Comp. Phys.} {\bf 82} 64
\item Bessel M S, Christlieb N and Gustafsson B 2004 {\it ApJL} {\bf 612} 61
\item Bogleux E and Galli D 1997 {\it MNRAS} {\bf 288} 638
\item Bond J R, Cole S, Efstathiou G and Kaiser N 1991 {\it ApJ} {\bf 379} 440
\item Bonnor W B 1956 {\it MNRAS} {\bf 116} 351
\item Bromm V, Coppi P and Larson R B 1999 {\it ApJL} {\bf 527} L5
\item Bromm V, Coppi P and Larson R B 2002 {\it ApJ} {\bf 564} 23
\item Bromm V, Ferrara A, Coppi P and Larson R B 2001 {\it MNRAS}
  {\bf 328} 969
\item Bromm V and Larson R B 2004 {\it ARAA} {\bf 42} 79
\item Bromm V, Kudritzki R P and Loeb A 2001 {\it ApJ} {\bf 552} 464
\item Bryan G L and Norman M 1998 {\it ApJ} {\bf 495} 80
\item Bryan G L, Norman M L, Stone J M, Cen R and Ostriker J P 1995 {\it
Comp. Phys. Comm.} {\bf 89} 149
\item Burkert A and Bodenheimer P 1993 {\it MNRAS} {\bf 264} 798
\item Cazaux S and Spaans M 2004 {\it ApJ} {\bf 611} 40
\item Cen R and Ostriker P 1992 {\it ApJL} {\bf 399} L113
\item Christlieb N, Bessel M S, Beers T C, Gustafsson B, Korn A, Barklem P S,
  Karlsson T, Mizuno-Wiedner M and Rossi S 2002 {\it Nature} {\bf 419} 904
\item Ciardi B and Ferrara A 2004 {\it Space Science Reviews} accepted
  [astro-ph/0409018]
\item Ciardi B, Ferrara Governato F  and Jenkins A {\it MNRAS} {\bf 314} 611   
\item Coles P and Lucchin F 1995 {\it Cosmology. The origin and
evolution of cosmic structure} (Chichester, England: John Wiley \& Sons)
\item Couchman H M and Rees M J 1986 {\it MNRAS} {\bf 221} 53
\item Cowie L L and Songaila A 1998 {\it Nature} {\bf 394} 44
\item Diemand J, Moore B and Stadel J 2005 {\it Nature} {\bf 433} 389 [astro-ph/0501589]
\item Doroshkevich A G, Zel'Dovich Y B and Novikov I D 1967
  {\it  Astronomicheskii Zhurnal} {\bf 44} 295 [DZN67]
\item Ebert R. 1955 {\it Z. Astrophys.} {\bf 37} 217
\item Fan X \etal 2000 {\it AJ} {\bf 120} 1167
\item Fan X \etal 2001 {\it AJ} {\bf 122} 2833
\item Fan X \etal 2003 {\it AJ} {\bf 125} 2151
\item Fan X \etal 2004 {\it AJ} {\bf 128} 1649
\item Ferrara A and Salvaterra R 2005 {\bf THIS BOOK}
  [see also astro-ph/0406554]
\item Ferrara A and Tolstoy E 2000 {\it MNRAS} {\bf 313} 291
\item Flower D R, Le Bourlot J, Pineau de For\^ets G and Roueff E 2000
  {\it MNRAS} {\bf 314} 753--758
\item Frommhold L 1993 {\it Collision-induced absorption in gases}
(Cambridge, UK: Cambridge University Press)
\item Fuller T M and Couchman H M P 2000 {\it ApJ} {\bf 544} 6--20
\item Galli D and Palla F 1998 {\it A\&A} {\bf 335} 403--420 [GP98]
\item Gamow G 1948 {\it Physical Review} {\bf 74} 505
\item Gamow G 1949 {\it Rev. Mod. Phys.} {\bf 21}, 367
\item Glover S C O 2004 {\it Space Science Reviews} accepted [astro-ph/0409737]
\item Glover S C O and Brand P W J L 2001 {\it MNRAS} {\bf 321} 385
\item Glover S C O and Brand P W J L 2003 {\it MNRAS} {\bf 340} 210
\item Gou L, Meszaros P, Abel T and Zhang B 2004 {\it Apj} {\bf 604} 508
\item Green A M, Hofmann S and Schwarz D J 2004 {\it MNRAS} {\bf 353} 23
\item Guth A H 2004 to be published in {\it Carnegie Observatories
  Astrophysics Series, Vol. 2: Measuring and Modeling the Universe},
  ed. W L Freedman (Cambridge: Cambridge University Press),
  [astro-ph/0404546]
\item Guth A H 1981 {\it Phys. Rev. D} {\bf 23} 247
\item Haardt F and Madau P 1996 {\it ApJ} {\bf 461} 20
\item Haiman Z, Abel T and Rees M J 2000 {\it ApJ} {\bf 534} 11
\item Haiman Z, Rees M J and Loeb A 1997 {\it ApJ} {\bf 476} 458
\item Heger A, Fryer C L, Woosley S E, Langer N and Hartmann D H 2003 {\it
  ApJ} {\bf 591} 288
\item Heger A and Woosley S E 2002 {\it ApJ} {\bf 567} 532
\item Heger A, Woosley S, Baraffe I and Abel T 2002 in {\it Lighthouses in
  the Universe; The Most Luminous Celestial Objects and Their Use for
  Cosmology}, proceedings of the MPA/ESO Conference held at Garching
  (Germany), eds. M. Gilfanov, R. A. Siunyaev and E. Curazov (Berlin:
  Springer), 369 [astro-ph/0112059]
\item Hollenbach D and McKee C F 1979 {\it ApJS} {\bf 41} 555--592
\item Kitayama T, Susa H, Umemura M and Ikeuchi S 2001 {\it MNRAS} {\bf 326} 1353
\item Larson R B 1969 {\it MNRAS} {\bf 145} 271
\item Le Bourlot J, Pineau de For\^ets G and Flower D R 1999
  {\it MNRAS} {\bf 305} 802
\item Lepp S and Shull J M 1983 {\it ApJ} {\bf 270} 578--582
\item Lepp S and Shull J M 1984 {\it ApJ} {\bf 280} 465--469
\item Lenzuni P, Chernoff D F and Salpeter E E 1991 {\it ApJS} {\bf 76} 759
\item Lynden-Bell D 1967 {\it MNRAS} {\bf 136} 101
\item Mac Low M M and Ferrara A 1999 {\it ApJ} {\bf 513} 142
\item Machacek M E, Bryan G L and Abel T 2001 {\it ApJ} {\bf 548}
  509--521
\item Madau P and Haardt F 2005 {\bf THIS BOOK}
\item Madau P, Ferrara A and Rees M J 2001 {\it ApJ} {\bf 555} 92
\item Madau P, Rees M J, Volonteri M, Haardt F and Oh S P {\it ApJ}
      {\bf 604} 484
\item Marri S and Ferrara A 1998 {\it ApJ} {\bf 509} 43
\item Marri S, Ferrara A and Pozzetti L 2000 {\it MNRAS} {\bf 317} 265
\item Martin P G, Schwarz D H and Mandy M E 1996 {\it ApJ} {\bf 461}
  265--281
\item McDowell M R C 1961 {\it Observatory} {\bf 81} 240
\item Monaghan J J 1992 {\it ARAA} {\bf 30} 543
\item Norman M L 2004 to be published in {\it Springer Lecture Notes in
Computational Science and Engineering: Adaptive Mesh Refinement -
Theory and Applications}, eds. T Plewa T Linde and G Weirs [astro-ph/0402230]
\item Novikov I D and Zel'Dovich Y B 1967 {\it ARAA} {\bf 5} 627
\item O'Shea B W, Bryan G, Bordner J, Norman M L, Abel T, Harkness R and
Kritsuk A 2004 to be published in {\it Springer Lecture Notes in
Computational Science and Engineering: Adaptive Mesh Refinement -
Theory and Applications}, eds. T Plewa T Linde and G Weirs [astro-ph/0403044]
\item Ober W W, El Elid M F and Fricke K J 1983 {\it A\&A} {\bf 119} 61
\item Omukai K 2001 {\it ApJ} {\bf 546} 635
\item Omukai K and Nishi R 1998 {\it ApJ} {\bf 508} 141--150
\item Omukai K and Palla F 2001 {\it ApJL} {\bf 561} L55--L58
\item Omukai K and Palla F 2003 {\it ApJ} {\bf 589} 677--687
\item Omukai K and Yoshii Y 2003 {\it ApJ} {\bf 599} 746
\item Osterbrock D E 1989 {\it Astrophysics of Gaseous Nebulae and
Active Galactic Nuclei} (Mill Valley, California: University Science Books)
\item Padmanabhan T 1993 {\it Structure Formation in the Universe}
  (Cambridge: Cambridge University Press)
\item Palla F, Salpeter E E and Stahler S W 1983 {\it ApJ} {\bf 271}
  632
\item Peebles P J E 1965 {\it ApJ} {\bf 142} 1317
\item Peebles P J E 1993 {\it Principles of Physical Cosmology} (Princeton: Princeton University Press)
\item Peebles P J E and Dicke R H 1968 {\it ApJ} {\bf 154} 891 [PD68]
\item Penston M V 1969 {\it MNRAS} {\bf 144} 425
\item Press W H and Schechter P 1974 {\it ApJ} {\bf 187} 425
\item Quinn P J and Zurek W H 1988 {\it ApJ} {\bf 331} 1
\item Rees M J 1976 {\it MNRAS} {\bf 174} 483
\item Ripamonti E and Abel T 2004 {\it MNRAS} {\bf 348} 1019--1034
\item Ripamonti E, Haardt F, Ferrara A and Colpi M 2002 {\it MNRAS}
  {\bf 334} 401--418
\item Sabano Y and Yoshii Y 1977 {\it PASJ} {\bf 29} 207
\item Saslaw W C and Zipoy D 1967 {\it Nature} {\bf 216} 976
\item Schaerer D 2002 {\it A\&A} {\bf 382} 28
\item Schneider R, Ferrara A, Natarajan P and Omukai K 2002 {\it ApJ}
  {\bf 571} 30
\item Shapiro P R, Iliev I T and Raga A C 2004 {\it MNRAS} {\bf 348} 753
\item Sigward F, Ferrara A and Scannapieco E 2004 {\it MNRAS}, accepted
   [astro-ph/0411187]
\item Silk J 1968 {\it ApJ} {\bf 151} 459--472
\item Silk J 1983 {\it MNRAS} {\bf 205} 705
\item Somerville R 2005 {\bf THIS BOOK}
\item Stahler S W, Shu F H and Taam R E 1980 {\it ApJ} {\bf 241} 637
\item Stahler S, Palla F and Salpeter E E 1986 {\it ApJ} {\bf 302} 590
\item Starobinsky A A 1979 {\it Pis'ma Zh. Eksp. Teor. Fiz.} {\bf 30}
  719 [JETP Lett. 30,682]
\item Starobinsky A A 1980 {\it Phys. Lett.} {\bf 91B} 99
\item Tegmark M, Silk J, Rees M J, Blanchard A, Abel T and Palla F 1997
  {\it ApJ} {\bf 474} 1--12 [T97]
\item Tumlinson J and Shull J M 2000 {\it ApJL} {\bf 528} 65
\item Uehara H and Inutsuka S 2000 {\it ApJL} {\bf 531} L91--L94 (HD)
\item Viana P T P and Liddle A R 1996 {\it MNRAS} {\bf 281} 323
\item Volonteri M, Haardt F and Madau P 2003 {\it ApJ} {\bf 582} 559
\item Walter F, Bertoldi F, Carilli C, Cox P, Lo K Y, Neri R, Fan X, Omont A,
      Strauss MA and Menten K M 2003 {\it Nature} {\bf 424} 406
\item Whalen D, Abel T and Norman M L 2004 {\it ApJ} {\bf 610} 14
\item White S M and Springel V 1999 {\it The First Stars}, proceedings
of the MPA/ESO Workshop held at Garching, Germany, 4-6 August 1999,
eds. A. Weiss, T. Abel \& V. Hill (Springer), p. 327 
\item Wise J H and Abel T 2004 {\it in preparation}, [astro-ph/0411558]
\item Woodward P R and Colella 1984 {\it J. Comp. Phys.} {\bf 54} 174
\item Wu K K S, Lahav O and Rees M J 1999 {\it Nature} {\bf 397} 225--30
\item Yoneyama T 1972 {\it Publ. of the Astr. Soc. of Japan} {\bf 24} 87
\item Yoshida N, Abel T, Hernquist L and Sugiyama N 2003 {\it ApJ}
  {\bf 592} 645--663
\item Zhao H S, Taylor J, Silk J and Hooper D 2005, astro-ph/0502049
\end{thereferences}

\end{document}